\pdfoutput=1 
\documentclass[a4paper,11pt]{article}
\usepackage[utf8]{inputenc}
\usepackage{jcappub} 
\usepackage{threeparttable}
\usepackage{tabularx}
\usepackage{multirow}
\usepackage{cleveref}
\usepackage{comment}
\usepackage{units}
\usepackage{subfigure}
\usepackage{amssymb}
\usepackage{amsmath}
\usepackage{verbatim}
\usepackage{multicol}
\usepackage{graphicx}
\usepackage{float}
\usepackage{hyperref}
\usepackage{upgreek}
\usepackage[dvipsnames]{xcolor}
\usepackage{lineno}
\usepackage{datetime}

\title{Spectroscopic analysis of the argon scintillation with a wavelength sensitive particle detector}

\author[a]{R.~Santorelli,}
\author[a]{E.~Sanchez Garcia,}
\author[a]{P.~Garcia~Abia,}
\author[b]{D.~Gonz\'alez-D\'iaz,}
\author[a]{R.~Lopez~Manzano,}
\author[a]{J.J.~Martinez~Morales,}  
\author[a]{V.~Pesudo}
\author[a]{and L.~Romero}

\affiliation[a]{CIEMAT, Div. de F{\'\i}sica de Particulas, Avda. Complutense, 40, Madrid, Spain}
\affiliation[b] {IGFAE, Campus Vida, R\'ua Xos\'e Mar\'ia Su\'arez N\'u\~nez, s/n, Santiago de Compostela,  Spain}

\emailAdd{roberto.santorelli@ciemat.es, edgar.sanchez@ciemat.es}

\abstract{
We performed a time-resolved spectroscopic study of the VUV/UV argon scintillation as a function of pressure and electric field, by means of a wavelength sensitive detector operated with different radioactive sources. 
Our work conveys new evidence of distinctive features of the argon light which are in contrast with the general assumption  that, for particle detection purposes,  the scintillation can be considered to be largely monochromatic at  128~nm (second continuum).
The wavelength and the time-resolved analysis of the photon emission reveal that the dominant component of the argon scintillation during first tens of ns is in the range [160,~325]~nm. This light  is  consistent with the third continuum emission from highly charged argon ions/molecules. This component of the scintillation is field-independent up to 25~V/cm/bar and shows a very mild dependence with pressure in the range [1,~16]~bar. The dynamics of the second continuum emission is dominated by the excimer formation time, whose variation as a function of the pressure has been measured. Additionally, the time and pressure-dependent features of electron-ion recombination, in the second continuum band, have been measured. This study opens new paths toward a novel particle identification technique based on the spectral information of the noble-elements scintillation light.
}

\usepackage[normalem]{ulem}

\definecolor{ao(english)}{rgb}{0.0, 0.5, 0.0}

\newcommand{\beq}{\begin{equation}}
\newcommand{\eeq}{\end{equation}}

\newcommand{\bear}{\begin{eqnarray}}
\newcommand{\eear}{\end{eqnarray}}

\newcommand{\tn}{\textnormal}

\begin{document}
\today ~~ \currenttime    

\maketitle



\section{Introduction}
\label{Sec:intro}

Over the last decade, argon and xenon detectors have attracted a lot of interest for their use in direct dark matter search \cite{DS50:2018ves,ArDM:2016hve,DEAP:2019imk,XE1T:2018,DS20:2017fik} and neutrino experiments \cite{EXO:2014awa,uBOONE:2016smi,DUNE:2017aow}, given their unique ionization and scintillation properties. In these experiments, the high-purity noble element bulk is used as target, as well as active and tracking medium for particles \cite{DUNE:HPGAR,Alvarez:2012sma}, attaining overall performances better than the ones reachable with other technologies.

A central aspect of the single or dual (gas-liquid) phase noble element detectors is the efficient collection and detection of the vacuum ultraviolet (VUV) scintillation photons, which provides calorimetric data, event time for the 3D event reconstruction, and particle identification capability~\cite{DEAP:2020zyy,ArDM:2017qre}. 
However, the details of the photon production mechanisms as well as the wavelength and temporal spectra are not fully understood. Experimental information, obtained largely from the closely related fields of photo-chemistry, plasma and laser physics, provides abundant evidence that the light emission mechanism  relies  on the bond created between excited and ground state atoms through 3-body collisions. As densities is similar or above those of noble gases at standard (STP) conditions, Ar$_{2}^{*}$ and Xe$_{2}^{*}$ \emph{excimers}, Rydberg states with a dimer core and a binding electron, form. Singlet and triplet excimer states undergo radiative de-excitation, giving rise to the so-called \emph{second excimer continuum}. This feature dominates the scintillation spectrum for gas pressures above 100~mbar, and results in relatively narrow emission bands ($\approx$ 10 nm wide) at 128~nm (argon)  and 172~nm (xenon). 

Under the above paradigm, there has been so far little motivation towards exploiting spectroscopic information in this kind of particle detectors. As a consequence, the light detection systems of argon and xenon chambers are based on broad-band optical sensors, possibly coupled to photon wavelength-shifters, that effectively integrate the light signal over a wide spectral range, missing the potential information provided by the scintillation wavelength. 

Some studies were carried out (mostly prior to the 90's) to investigate the temporal evolution of characteristic spectral components of the argon and xenon scintillation. This experimental program was performed through irradiation with strong X-ray \cite{xrays}, electron \cite{Koehler, Langhoff_e}, proton \cite{Hurst} or heavy ion \cite{heavy1, Wiese} beams.   Although  a high intensity beam can produce, in a relatively simple way, the photon yield required for the  study of scintillation in pure gases, spurious effects (associative processes, interactions with walls and emission from the window) are often introduced. 
Comparatively, it is more difficult to study the time-resolved spectroscopy with individual particles, and, for this reason, the systematic investigation of the spectral characteristics and absolute scintillation yields in particle detectors is much scarcer, and seems to be almost entirely circumscribed to $\upalpha$-particles \cite{Carvalho, Birot:1975, Suzuki:1981}.

With regard to the particle detectors field, the demonstration of a light emission mechanism more complex than the one so far considered in the literature, with well separated emission bands for different particle types, could trigger the interest on a novel detector concept sensitive to the photon wavelength. This technology would allow to exploit distinctive features of the noble gases scintillation emission, a possibility currently not conceived by the present experiments. 

To this aim, we have built a high-pressure gas detector in the CIEMAT laboratory (Sec.~\ref{Sec:ES}). The use of photon sensors with different spectral sensitivities allowed us to detect the Ar scintillation  in the near, middle and far UV ranges, and to investigate, in detail, aspects of the photon production mechanism and light emission spectrum particularly relevant for particle detectors based on noble elements. 

Results obtained with $\upalpha$ and $\upbeta$ sources deployed in the gas chamber, operated up to 20 bar, evidenced a substantial emission in the middle-UV region, which is consistent with the so-called \textit{third continuum emission} (Secs.~\ref{Sec:sci} and~\ref{Sec:ResSP}). Despite being still sub-dominant, overall, compared to the second continuum, the third continuum is markedly fast, hence representing the main contribution to the photon signal during the first tens of ns (Secs.~\ref{Sec:ResSP} and~\ref{Sec:ResP}) below 10 bar. 

To the best of our knowledge, this is the first systematic study of the third continuum carried out with $\upalpha$ and $\upbeta$ sources  operated in a noble-gas chamber. Our findings significantly boost the importance of the spectral information in noble-element detectors. Although the two continua can be unambiguously differentiated spectroscopically, the  widespread use of wavelength shifter coatings in dark matter and neutrino detectors spoils any possibility of exploiting the spectroscopic information of the argon scintillation.

\section{Brief review of the scintillation in noble gases}
\label{Sec:sci}

In this section we revisit briefly the experimental situation, starting with the $1^{\tn{st}}$ and $2^{\tn{nd}}$ continuum, that have their origin on the emission from vibrationally-relaxed singlet ($\tn{Ar}_2^*(^1\Sigma_u)$) and triplet ($\tn{Ar}_2^*(^3\Sigma_u)$) excimer states, when transitioning to the dissociative  ground state of the Ar dimer ($\tn{Ar}_2(^1\Sigma_g)$). The importance of this emission stems from the fact that, for pressures above few 100’s of mbar, low-lying excited atomic states (resonant: $\tn{Ar}^*(s_4)$, metastable: $\tn{Ar}^*(s_5)$) are quickly and predominantly populated through a collision-dominated atomic cascade; given that singlet and triplet excimer states are formed from those atomic states through termolecular (3-body) reactions, this emission displays a high universality. Focusing on the $\tn{Ar}^*(s_4)$ state, the scintillation process can be characterized in general through the following (dominant) pathways:
\bear
\tn{Ar}^{**} + \tn{Ar}                &\xrightarrow{\tn{(cascade)}} ~~\tn{Ar}^*(s_4) + \tn{Ar} ~~~~~~~~~~~~~~~~&\tn{collisional quenching} \label{argon1}\\
\tn{Ar}^{**}                &\xrightarrow{\tn{(cascade)}} ~~ \tn{Ar}^*(s_4) + h\nu\tn{'s} ~~~~~~~~~~~~~~&\tn{atomic radiation} \label{argon2}\\
\tn{Ar}^*(s_4) + \tn{Ar} + \tn{Ar}         &\xrightarrow{K_{s_4\rightarrow{0_u}}} ~~\tn{Ar}_2^*(0_u^+) + \tn{Ar} ~~~~~~~~~~~~~~~&\tn{(weakly bound) exc. form.} \\
\tn{Ar}_2^*(0_u^+)                 &\xrightarrow{K_{0_u\rightarrow{s_4}}}   ~~\tn{Ar}^*(s_4) + \tn{Ar} ~~~~~~~~~~~~~~~~ &\tn{exc. dissociation to} ~ s_4 \\
\tn{Ar}_2^*(0_u^+)                 &\xrightarrow{K_{0_u\rightarrow{s_5}}}   ~~\tn{Ar}^*(s_5) + \tn{Ar} ~~~~~~~~~~~~~~~~&\tn{exc. dissociation to} ~ s_5 \\
\tn{Ar}_2^*(0_u^+)                 &\xrightarrow{1/\tau_{_{0_u}}} ~~~  \tn{Ar} + \tn{Ar} + h\nu ~~~~~~~~~~~~~~&\tn{first continuum} \\
\tn{Ar}_2^*(0_u^+) + \tn{Ar}        &\xrightarrow{K_{bind,1}}   ~~\tn{Ar}_2^*(^1\Sigma_u(v)) + \tn{Ar} ~~~~~~~~~~&\tn{exc. binding} \label{NONAME} \\
\tn{Ar}_2^*(^1\Sigma_u(v)) + \tn{Ar}  &\xrightarrow{K_{cool,1}}   ~~\tn{Ar}_2^*(^1\Sigma_u(v=0)) + \tn{Ar} ~~~~&\tn{exc. cooling/relaxation} \\
\tn{Ar}_2^*(^1\Sigma_u(v=0))     &\xrightarrow{1/\tau_{e_{2}}^{s}}  ~~~~\tn{Ar}_2(^1\Sigma_g) + \tn{Ar} + h\nu  ~~~~~~ &\tn{second continuum}\label{argonlast}
\eear
Here $\tn{Ar}^{**}$ refers to any excited atomic state above the two lowest ones,  $\tn{Ar}_2^*(0_u^+)$ is the weakly bound $\tn{Ar}_2^*$ excimer associated with $\tn{Ar}_2^*(^1\Sigma_u(v))$, in Hund's notation corresponding to \emph{case c}, and the quantum number $v$ refers to the vibrational state, with $v=0$ corresponding to the bottom of the potential well. If considering, instead, that the cascade proceeds through the $\tn{Ar}^*(s_5)$ state, the situation is analogous:
\bear
\tn{Ar}^{**} + \tn{Ar}                &\xrightarrow{\tn{(cascade)}} ~~\tn{Ar}^*(s_5) + \tn{Ar} ~~~~~~~~~~~~~~~~&\tn{collisional quenching} \label{argon1_}\\
\tn{Ar}^{**}                &\xrightarrow{\tn{(cascade)}} ~~ \tn{Ar}^*(s_5) + h\nu\tn{'s} ~~~~~~~~~~~~~~&\tn{atomic radiation} \label{argon2_}\\
\tn{Ar}^*(s_5) + \tn{Ar} + \tn{Ar}         &\xrightarrow{K_{s_5\rightarrow{0_u}}} ~~\tn{Ar}_2^*(1_u/0_u^-) + \tn{Ar} ~~~~~~~~~~~&\tn{(weakly bound) exc. form.~~~~} \\
\tn{Ar}_2^*(1_u/0_u^-)                 &\xrightarrow{K_{0_u\rightarrow{s_5}}}   ~~~\tn{Ar}^*(s_5) + \tn{Ar} ~~~~~~~~~~~~~~~~ &\tn{dissociation to} ~ s_5 \\
\tn{Ar}_2^*(1_u/0_u^-)                 &\xrightarrow{1/\tau_{_{1_u}}} ~~~~\tn{Ar} + \tn{Ar} + h\nu ~~~~~~~~~~~~~~~ &\tn{first continuum} \\
\tn{Ar}_2^*(1_u/0_u^-) + \tn{Ar}        &\xrightarrow{K_{bind,3}}   ~~\tn{Ar}_2^*(^3\Sigma_u(v)) + \tn{Ar} ~~~~~~~~~~&\tn{exc. binding} \label{NONAME_} \\
\tn{Ar}_2^*(^3\Sigma_u(v)) + \tn{Ar}  &\xrightarrow{K_{cool,3}}   ~~\tn{Ar}_2^*(^3\Sigma_u(v=0)) + \tn{Ar} ~~~~&\tn{exc. cooling/relaxation} \\
\tn{Ar}_2^*(^3\Sigma_u(v=0))     &\xrightarrow{1/\tau_{e_{2}}^{t}}   ~~~~\tn{Ar}_2(^1\Sigma_g) + \tn{Ar} + h\nu  ~~~~~~ &\tn{second continuum}\label{argonlast_}
\eear
and $\tn{Ar}_2^*(1_u/0_u^-)$ refers to the degenerate weakly bound excimers associated with $\tn{Ar}_2^*(^3\Sigma_u(v))$. A detailed diagram compiling these pathways can be found for instance in \cite{DiegoMicro}.  Reaction rates ($K$) will be assumed in this work to be in units of  [t$^{-1}$], therefore being number density ($N$) dependent. Experimentally, it is has been determined that, for pressures above 100~mbar, the dominant time constants are the (3-body) formation times $\tau_{f_{2}} = 1/K_{s_{5(4)} \rightarrow 0_u}$, and the decay times $\tau_{e_{2}}^{s}$, $\tau_{e_{2}}^{t}$. 

At high ionization densities, characteristic of high pressures and/or highly ionizing radiation, charge recombination provides additional scintillation channels \cite{RECO1, RECO2}. From
\bear
\tn{Ar}^{+} + \tn{Ar} + \tn{Ar} &\rightarrow& \tn{Ar}_2^+ + \tn{Ar} \\
\tn{Ar}_2^{+} + e &\rightarrow& \tn{Ar}_2^{**} \label{RECO_e}\\
\tn{Ar}_2^{**} &\rightarrow& \tn{Ar}^{**} + \tn{Ar}
\eear
the $1^{\tn{st}}$ and $2^{\tn{nd}}$ continua follow, according to reactions \ref{argon1}, \ref{argon2}, \ref{argon1_}, \ref{argon2_}. This leads to a perfect correlation between recombined charge and excess scintillation, with reaction \ref{RECO_e} being regulated by the external electric field \cite{SuzuField}. As a result, the effective energy to create a UV photon at zero-field varies, under $\upalpha$-particles, from $W_{sc}$=50.6~eV at 2~bar to $W_{sc}$=25.3~eV at 10~bar \cite{Saito}.

On the other hand, mechanisms leading to scintillation mainly in the region [180,~300]~nm are grouped under the generic denomination of ``$3^{rd}$ continuum'', despite it is possible that a number of different species contribute, as hinted by the study in \cite{Wiese}. The identification of the precursors to this continuum has been subject of hot debates throughout the 80's and 90's \cite{Langhoff:1994, Boisenko}. The most complete study to date, combining the two leading hypothesis, doubly ionized (Ar$^{++}$) and excited (Ar$^{+*}$) ions, seems to be \cite{Wiese}, however it does not allow quantitative predictions of the scintillation yields, nor it has been fully substantiated. 
In a nutshell, doubly ionized states would  lead to scintillation around 200~nm through three-body reactions, as:
\bear
\tn{Ar}^{++} + \tn{Ar} + \tn{Ar} & \rightarrow & (\tn{Ar}_2)^{++}   +   \tn{Ar} \label{3rd_ch1}\\ 
(\tn{Ar}_2)^{++}	  & \rightarrow & \tn{Ar}^+   + \tn{Ar}^+    +    h\nu
\eear
At high pressures, the presence of new peaks in the range [200,~300]~nm led to the consideration of additional contributions, finding a good theoretical correspondence with $\tn{Ar}^{+*}$ decays:
\bear
\tn{Ar}^{+*}	 & \rightarrow & \tn{Ar}^{+}   +   h\nu
\eear
Contrary to the $1^{\tn{st}}$ and $2^{\tn{nd}}$ continua, the pathways leading to formation of $\tn{Ar}^{+*}$ species have not been unambiguously identified. 
According to experimental data obtained in \cite{Carvalho} for $\upalpha$-particles, electron-ion recombination is not competing with process \ref{3rd_ch1}, so the $3^{\tn{rd}}$ continuum would be largely field-independent except perhaps for very highly ionizing radiation.

\section{Experimental setup} 
\label{Sec:ES}

\begin{figure}[!t]
\centering
\includegraphics[width=0.5\textwidth]{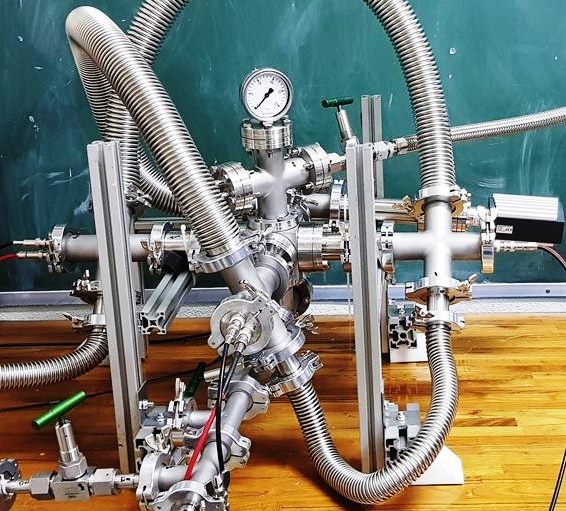}
\includegraphics[width=0.48\textwidth]{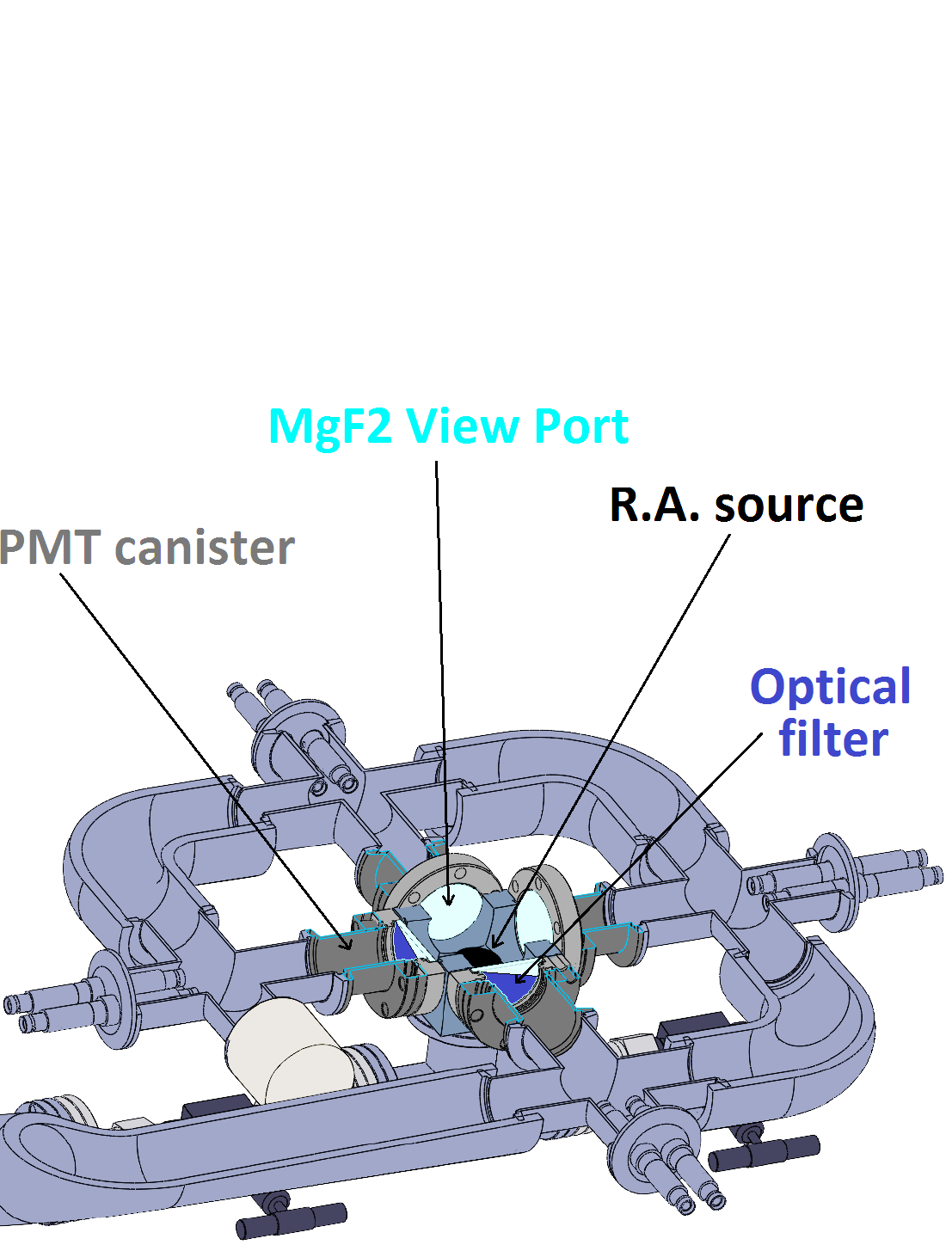}
\caption{(Left) Photograph of the experimental setup. (Right) Top view of detector. The pressurized region with the radioactive source is in the centre of the chamber. }
\label{detector}
\end{figure}

\begin{figure}[!t]
\centering
\includegraphics[width=0.49\textwidth]{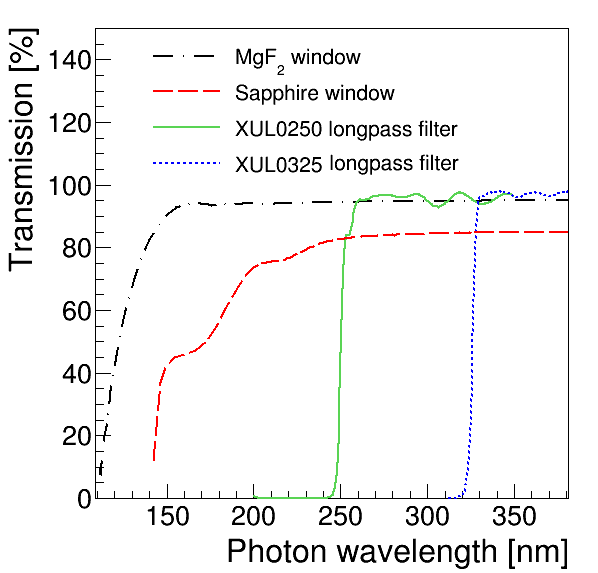}
\hspace{0.3 pt}
\includegraphics[width=0.49\textwidth]{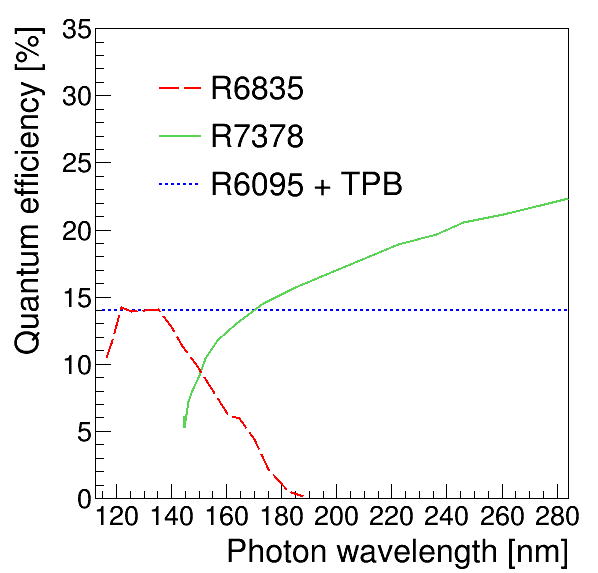}
\caption[]{(Left) Curves of light transmission as a function of wavelength, for the windows and filters used in the experiment. (Right) Quantum efficiencies of the PMTs used in the detector, measured by Hamamatsu as a function of the wavelength of the incident photons.}
\label{efficiency}
\end{figure}

The wavelength-sensitive chamber is shown in Fig \ref{detector}. It consists of a 7 cm side stainless steel cube with one CF-40 flange on each face. The top and bottom flanges are connected to the service lines for the vacuum and the gas filling. The lateral flanges are equipped with four custom-made optical view-ports necessary to decouple the pressurized region from the optical readout system.  A custom-designed steel frame keeps the windows in place. Grooves in the mechanical assembly hold viton o-rings which seal the window, making the view-port vacuum and pressure-tight up to 20 bar. 
The windows are made of  8.0 mm thick UV-grade MgF$_2$ crystals (38.1~mm--$\varnothing$), whose transmission is $\approx$~95\% for wavelengths above 180~nm (Fig.~\ref{efficiency}-left). A transmission of $\approx$~33\% at 128~nm is measured comparing the  light detected with and without the  MgF$_2$ crystals installed in the chamber.

Four 1" Hamamatsu photomultipliers (PMT) are hosted in canisters attached to the optical view-ports. They are built with KF-40 tee and cross fittings and equipped with a connection for the vacuum line, separated from the main pressure chamber, and two electrical feedthroughs for the HV and  PMT signals.

PMTs with different spectral sensitivities are used in the experiment. Two Hamamatsu R7378A phototubes are placed in opposite sides of the central chamber. In the other two confronting faces, one Hamamatsu R6095 and one Hamamatsu R6835 are set. The Hamamatsu R6095 PMT is coated with a 200~$\upmu$g/cm$^{2}$  layer of tetraphenyl butadiene (TPB), a wavelength shifter that  converts the UV photons to 420~nm with approximately 100~\% efficiency~\cite{Benson:2018}. The nominal quantum efficiencies (QE) of these phototubes and the MgF$_2$ window's transmission coefficient are plotted in Fig.~\ref{efficiency}-right as a function of the photon wavelength~\cite{Hamamatsu:R7378,Hamamatsu:R6835}.

The  TPB  coating  makes the R6095 response independent of the wavelength of the incident photon in the VUV-UV region, so its spectral sensitivity is considered constant in the range of interest ($[100,~300]$~nm). The QE of this tube is considered equal to half of the nominal value at 420~nm (30 \%) to take into account the 50 \% probability of backward emission of the TPB-converted photons. The CsI photocathode of the R6835 is solar blind, thus the 420 nm  photons, diffused backwards by the TPB coating of the R6095 placed in front of it, do not produce a signal. Specific tests performed in our laboratory with the R6835 PMT and a blue LED proved that this phototube is not sensitive to the 420~nm light. The geometry of the setup reduces  the optical cross-talk between the 420~nm photons and the R7378 phototubes to a level which is negligible for our purposes.

Special runs with  1" ($\varnothing$)  Al$_{2}$O$_{3}$ (Sapphire) crystals from Thorlabs~\cite{Thorlabs:Sapp} and long-pass filters form Asahi~\cite{Asahi:XUL0250}),  placed in front of the R6835 and R7378 respectively, are also taken in order to narrow the spectral sensitivity of these phototubes. The cross-comparison of the signals with and without the filters placed in front of the PMTs allows to study the Ar scintillation in well defined UV ranges. This work investigates the argon scintillation in the spectral ranges $[110,~140]$~nm (called UV2 region in the following) and $[160,~325]$~nm (UV3 region) using the signals of the R6835 PMT with the Al$_{2}$O$_{3}$ window and of the R7378 PMT with the Asahi filters, respectively (Table~\ref{tab:Rangedef}). The upper limits are identified by the fact that no light signal is detected with the filters placed in front of the PMTs.

The detector is equipped with two pressure sensors (ITR-90) and a set of  1/2" VCR valves that allow to set the pressure in the chamber and the optical readout system independently. The optical filters can be put or removed without the need to open or evacuate the pressure chamber. Before each run, the system is pumped with  a TURBOVAC 350i for 24 hours and a pressure of  $2\times10^{-5}$~mbar is typically reached in the central volume.  Pure argon gas (AirLiquide ALPHAGAZ\texttrademark-2, purity $\ge$ 99.9999 \%) is used to fill the chamber. The PMT region is evacuated and continuously pumped during the data taking (pressure $<$9$\times$10$^{-5}$ mbar) to minimize the absorption of the UV-VUV photons by the air between the optical view-port and the PMT window.

The PMTs are powered independently with a negative bias voltage through custom-made bleeder circuits with SMD components mounted on PCBs. The signal is amplified 10 times with a CAEN N979 module. One of the two outputs of the fast amplifier goes to a leading edge discriminator (CAEN N841), with a threshold set to 0.5 photoelectrons. Different trigger logics (CAEN N455 quad logic unit), based on the coincidence of the R6095 and R7378 phototubes, are used  in the experiment. The second output of the fast amplifier is fed to a CAEN DT5725 sampling ADC (14-bit resolution and 250 MS/s sampling rate). Different runs are taken with digitization windows from 4~$\upmu$s to 16~$\upmu$s. The data are transferred via optical link to a computer for the analysis. The gain of the PMTs is obtained from dark-rate data taken in vacuum, by fitting the  single photoelectron peak with a Gaussian function. The calibration of the R6835 is performed directly with the tail of the argon light pulse, given its very low dark pulse rate. The high voltage of the PMTs is set independently, in order to equalize the gains to the level of 500~ADC counts per photoelectron.

\begin{center}
\begin{table}[t!]
\begin{tabular}{ c | c | c | c }
\hline
Region & Range &  Lower bound & Upper bound \\ 
\hline 
\hline 
\rule{0pt}{3.0ex}
  UV2 & $[110,~140]$ nm & R6835 lower spectral sensitivity limit &  R6835 + Al$_{2}$O$_{3}$ crystal\\ 
 \rule{0pt}{3.0ex}
 UV3 & $[160,~325]$ nm & R7378 lower spectral sensitivity limit & R7378 + XUL0325 filter \\ \hline
\end{tabular}
\caption{Definition of the UV2 and UV3 spectral ranges. The lower bounds are set by the spectral sensitivity of the PMTs. The upper limits are defined from the filters' transmission edge.}
\label{tab:Rangedef}
\end{table}
\end{center}

The gas purity is assessed through the decay time of the slow component of the argon second continuum emission. Values of $\approx$~3 $\upmu$s were obtained depending on the gas pressure and flow. Simple selection criteria are applied to reject a few percent of the total triggers which are produced by electronic noise and cross-talk events.

The results presented in this study are obtained with $\rm ^{241}$Am (activity $\approx$~500~Bq) and $\rm ^{90}Sr/Y$ (activity $\approx$~100~Bq) radioactive sources electrodeposited on stainless steel disks. The pressure chamber is equipped with a PTFE support structure able to host one radioactive source in the centre. 

\section{Wavelength-resolved UV/VUV emission with a $^{241}$Am source}
\label{Sec:ResSP}

\begin{figure}[!t]
\centering
\includegraphics[width=0.49\textwidth]{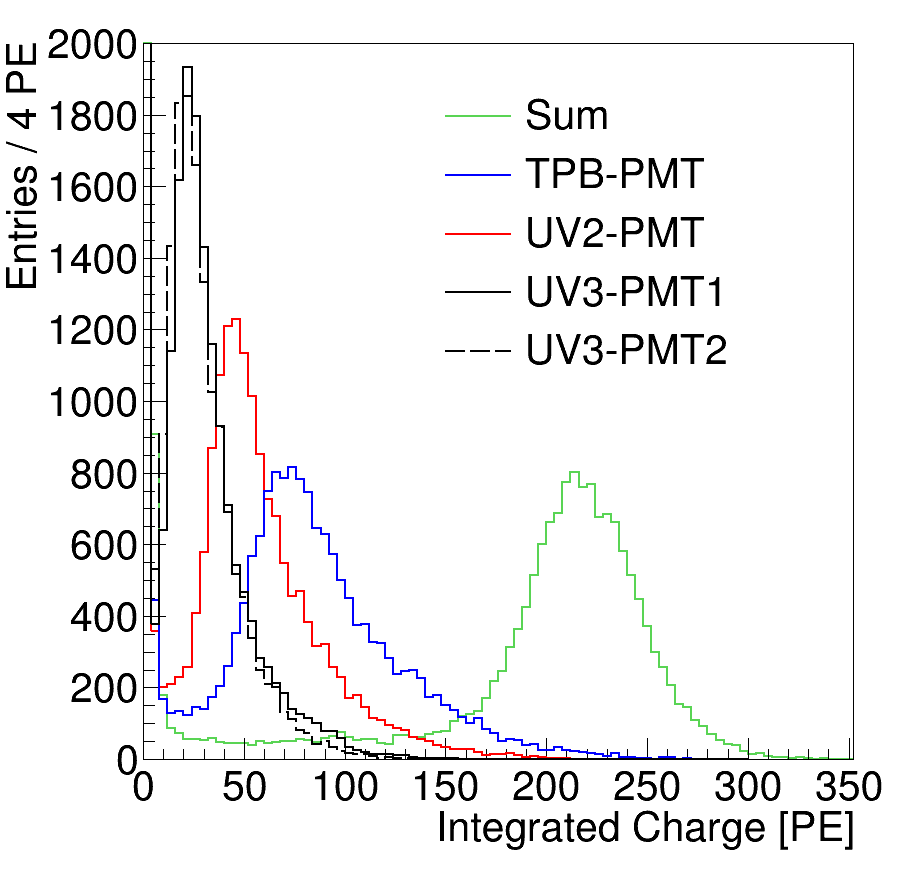}
\hspace{0.001 pt}
\includegraphics[width=0.49\textwidth]{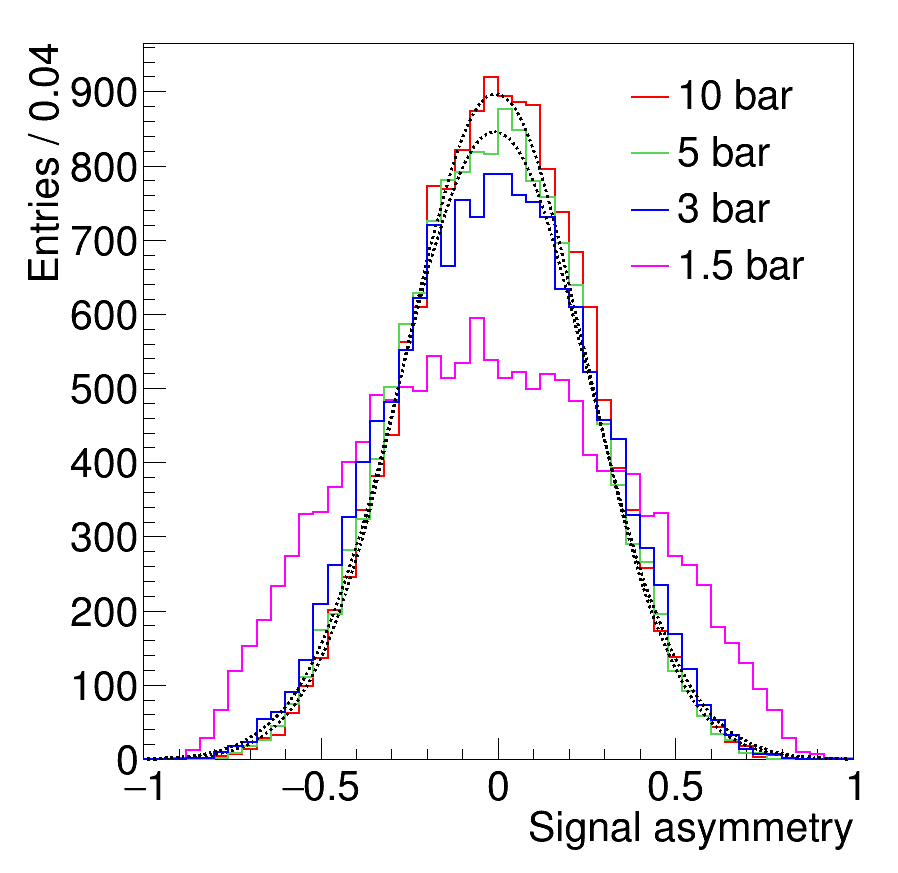}
\caption{(Left) Integrated charge signal of the four different PMTs used in the experiment, obtained with the $\rm ^{241}$Am $\upalpha$ source in argon at 2 bar. The signal in the UV3 region is detected with two R7378 PMTs (UV3-PMT1 and UV3-PMT2). The light collected on the four photo detectors (Sum) is around 220~PE. (Right) Asymmetry between the two R7378 PMTs for different pressures. The Gaussian functions fitted to the data are overlaid.}
\label{source_evidence}
\end{figure}

Initial measurements were carried out with a $\rm ^{241}$Am $\upalpha$-source in argon gas up to 16~bar using two R7378 (UV3-PMT1 and UV3-PMT2 in the following), one R6835 (UV2-PMT) and one TPB-coated R6095 PMT (TPB-PMT). Taking into account the W-values reported in \cite{Saito}, 1.1$\times$ 10$^{6}$ photons are produced in argon at 2 bar. Considering the solid angle between the PMTs photocathode and the source, the MgF$_2$ transmission coefficient and the nominal PMTs QE, (1.0~$\pm$~0.2) $\times$ 10$^{2}$ photoelectrons (PE) are expected to be detected by the  TPB-PMT, which is sensitive to the entire spectral range. The average solid angle between the PMTs photocathode and the source is estimated by means of a toy Monte Carlo which includes the detailed description of the chamber geometry and takes into account the refraction of  the light on the MgF$_2$ crystals. Values in the range [0.008,~0.011] are obtained for the different PMTs depending on the nominal photocathode size and the average length of the alpha track as a function of the gas pressure.

The $\upalpha$-peak is clearly visible in the integrated charge spectra (Fig.~\ref{source_evidence}-left). Depending on the gas purity and pressure, more than two hundreds of photoelectrons (PE) are detected in the full energy peak (green  histogram) produced by the $\approx$~5.5 MeV $\upalpha$ particles, allowing wavelength and time-resolved analysis of the light pulses. The TPB-PMT detects $\approx$~90~PE on average (blue  histogram in Fig.~\ref{source_evidence}-left), a value which is in  good agreement with the expected one.   The UV2-PMT (red histogram) mean value is approximately 60~PE, while the two UV3-PMTs (black histograms), which are only sensitive to wavelengths above 160~nm, detect $\approx$~30 PE each on average. This clearly establishes that there is a significant component of the argon scintillation light at wavelengths significantly longer than the 128~nm line from the second continuum. 

\begin{figure*}[!t]
\centering
\includegraphics[scale=0.375]{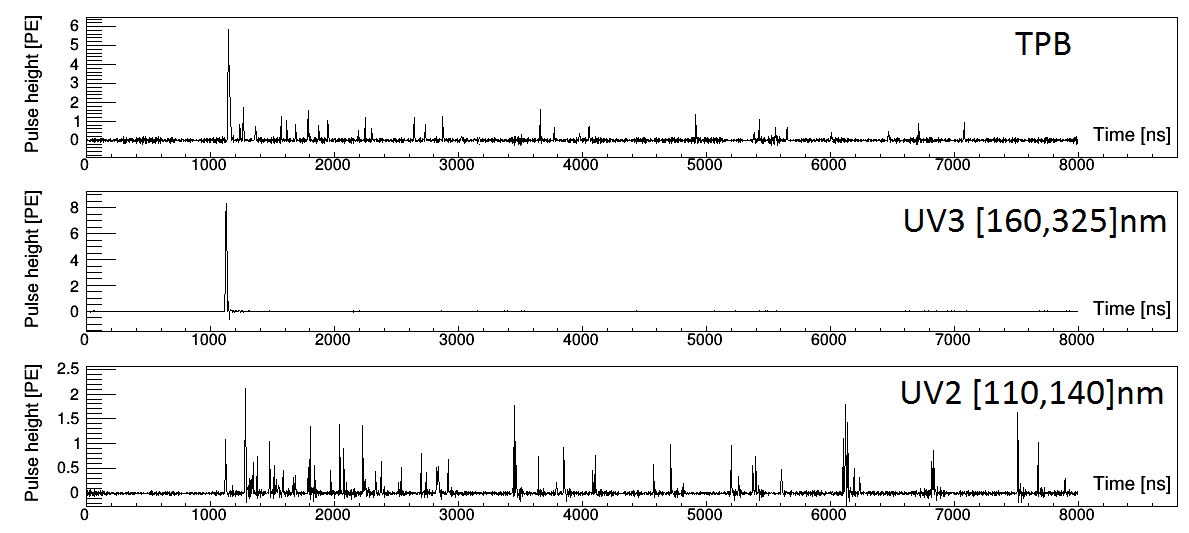}\par
\caption{Typical Ar scintillation signals from 5.5~MeV $\upalpha$-particle interaction  detected in three spectral regions, with argon gas at 1.5~bar. A clear separation of the components is possible by wavelength discrimination.}
\label{pulse_1bar}
\end{figure*}

The signal asymmetry, defined as the ratio between the difference and the sum of a pair of PMTs integrated signals, is plotted in Fig.~\ref{source_evidence}-right for the two UV3 phototubes at different pressure values. The data are fitted with a Gaussian function with a mean value compatible with zero, confirming that the source is at the centre of the chamber. The width of the distribution decreases with pressure due to the reduction of the mean range of the particles in the gas, indicating that the influence of any geometrical correction in the measured scintillation is  negligible above 3 bar.

The pulse shape of a typical signal, produced by an $\upalpha$ event in argon at 1.5 bar, is presented in Fig.~\ref{pulse_1bar}, as registered by the different PMTs. The TPB-PMT detects the characteristic fast and slow components of the argon emission. However, only the slow component of the scintillation is detected in the UV2 region, in form of a long train of pulses ($\mu$s scale) at the single photoelectron level. The UV3 sensitive phototubes are able to detect only the prompt emission, with a pulse amplitude similar to the one detected by the TPB-PMT.

A signal consistent with dark current is registered by the UV2-PMT with the sapphire crystal placed between the PMT window and the MgF$_{2}$ optical view-port, proving that the slow component of the Ar scintillation is entirely in the range $[110,~140]$~nm. The light signal detected by the UV3-PMTs with the 250~nm long-pass filter is of the order of 20~\% of the total signal without filter. No light is detected with the 325~nm long-pass filter. 

\begin{figure}[!t]
\centering
\includegraphics[width=0.49\textwidth]{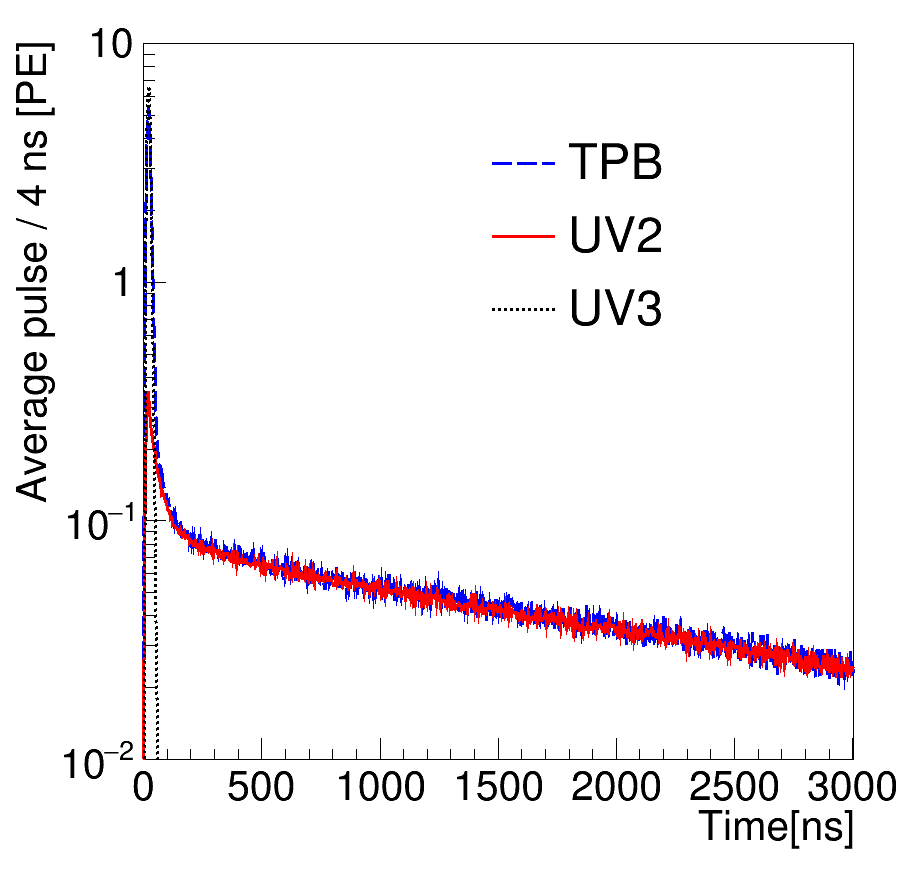}
\hspace{0.3 pt}
\includegraphics[width=0.49\textwidth]{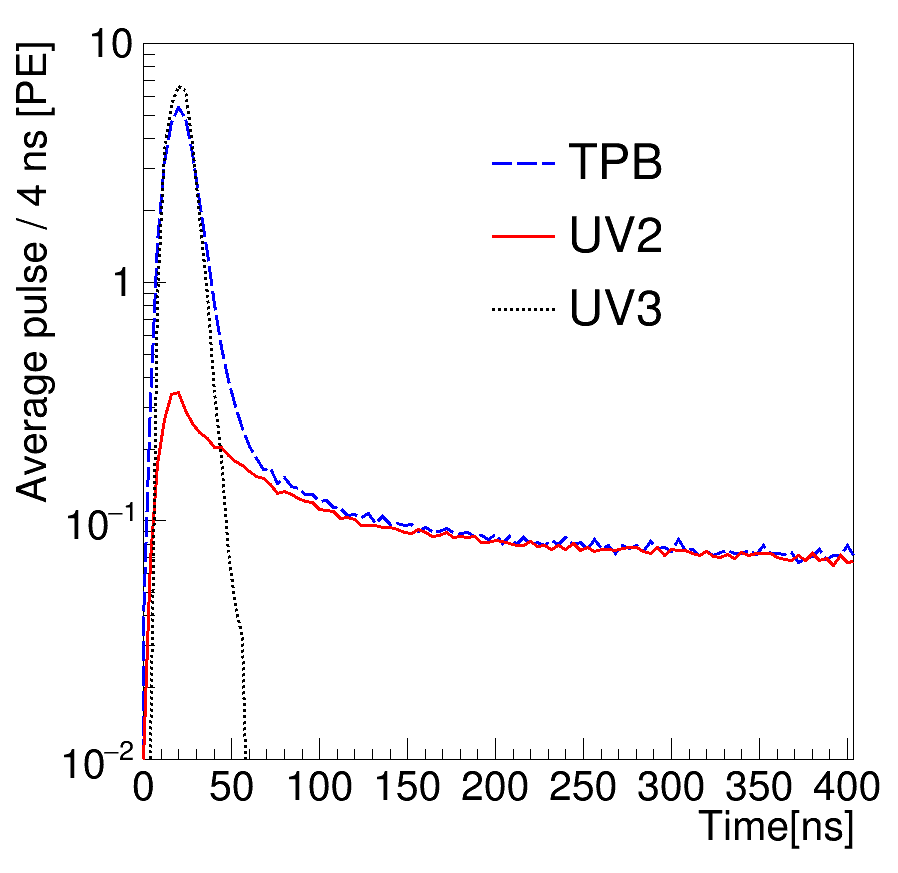}
\caption[]{Pulse shapes of the argon scintillation signals averaged over $3\times 10^4$ events per PMT type, registered with the $\rm ^{241}$Am source in Ar at 1.5 bar, in two time windows: (left) $[0, 3000]$~ns and (right) $[0,~400]$~ns.}
\label{argon_mean}
\end{figure}

The 1.5 bar argon scintillation pulses, averaged over $3\times 10^4$~events, are displayed in Fig.~\ref{argon_mean} in two time ranges. Only the $\upalpha$-particle interactions from the $\rm ^{241}$Am source are considered, using an energy threshold cut. LED calibration runs taken in vacuum evidenced an after-pulsing component, between 60 ns and 100 ns after the maximum pulse height, in the UV3-PMTs signal. This second pulse is removed from the analysis with a software cut, after confirming that the impact in our study is negligible.

\begin{figure}[!t]
\centering
\includegraphics[width=0.49\textwidth]{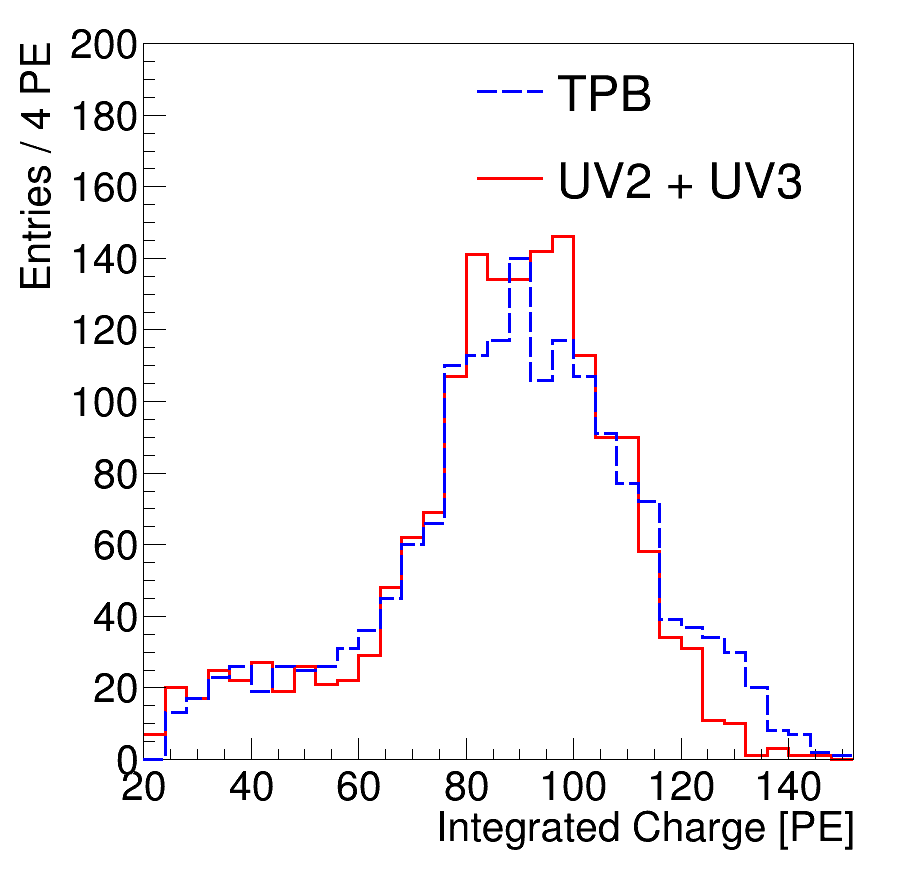}
\hspace{0.3 pt}
\includegraphics[width=0.49\textwidth]{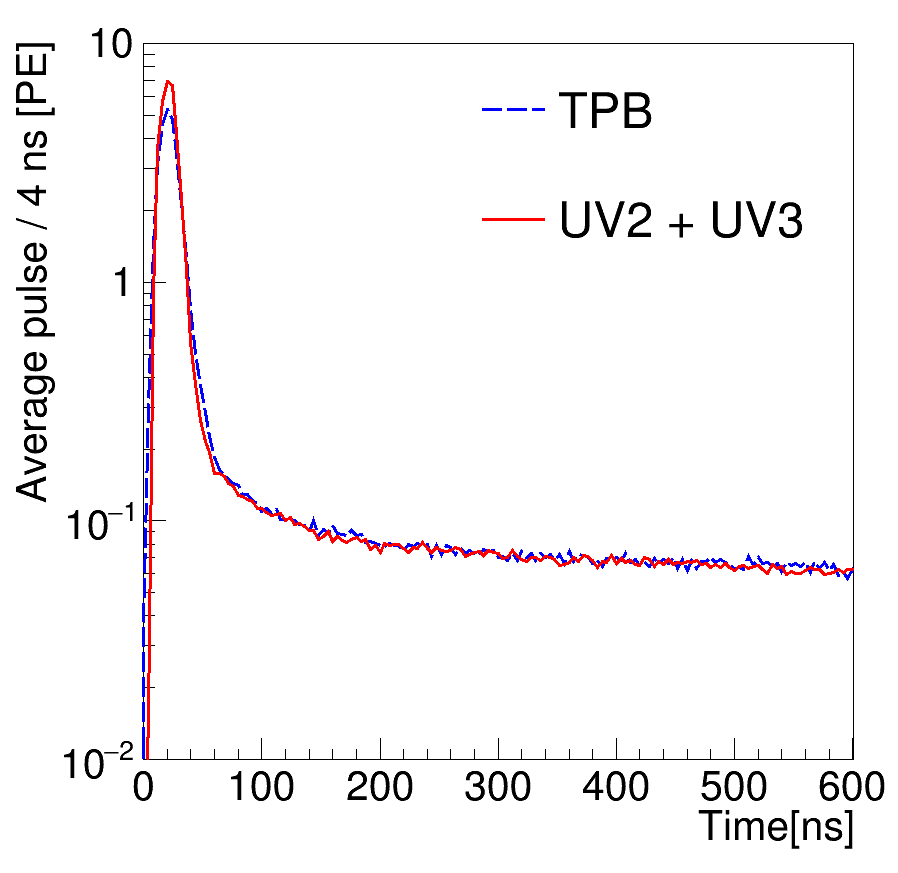}
\caption{(Left) Integrated charge spectrum obtained in the PMT with the wavelength shifter compared with the sum of the spectra in the UV2 and UV3 regions, with $\rm ^{241}$Am data in argon gas at 1.5 bar. (Right) Average argon pulse detected with the TPB and sum of the UV2 and UV3 pulses at 1.5~bar.}
\label{sum_comparison}
\end{figure}

In this analysis, only the raw number of PEs have been reported without any signal correction. The combination of the plots in Figs.~\ref{source_evidence}~(left) and~\ref{argon_mean} suggests that the amplitude of the signal detected by the TPB-PMT, which is sensitive to the entire spectral range, is very similar to the sum of the signals of two other PMTs, sensitive only to either of the two UV2 and UV3 regions. In Fig.~\ref{sum_comparison}, the integrated charge spectra (left) and the  average pulse shape of the argon scintillation (right) are shown with the $^{241}$Am source in argon gas at 1.5~bar, selecting the events in the center of the detector through two symmetry cuts between the different couples of confronting PMTs. Given the symmetry of the setup and the similar size of the PMTs, the fact that the tube signals are similar in the two regions UV2 and UV3 respectively  also suggests that the QEs are comparable.

These results prove that the two components of the scintillation with Ar gas at  1.5 bar are characterized by distinct wavelengths and can be unambiguously differentiated spectroscopically. The slow component is consistent with the second  continuum emission (Sec. \ref{Sec:sci}), which gives rise to the 128~nm photons. The fast component of the scintillation, on the other hand, is entirely in the UV3 region and it is compatible with the third continuum emission.

In our experiment the actual purity of the argon gas was estimated through the decay time constant of the triplet  excimer states $\tn{Ar}_2^*(^3\Sigma_u)$. The values obtained between 2.5 and 3 $\upmu$s set the  level of impurities in the range 0.1--1~ppm \cite{Akashi-Ronquest:2019mlk}, which is consistent with the gas contamination certified by the producer. The measured triplet decay time rejects the possible explanation of the UV3-photon production in terms of parasitic re-emissions from N$_{2}$, H$_{2}$O and O$_{2}$ contamination, which can only produce a sub-dominant component in the UV band and over a wide time scale. Any explanation, other than the third continuum emission, in terms of unusual Ar contaminants is considered implausible. 

The fast/slow component ratio obtained at 1.5~bar gas pressure and maximum purity (3.2~$\upmu$s) is 5.4, in good agreement with the value 5.5~$\pm$~0.6 at 1.1 bar absolute pressure measured in \cite{Amsler:2007gs}.

\section{Spectroscopic analysis of the argon scintillation as a function of the gas pressure }
\label{Sec:ResP}

In order to quantitatively compare signals in the two spectral regions UV2 and UV3, we calculate the total numbers of photons produced by the $\upalpha$ interactions considering the MgF$_{2}$  transmission values (33\% for the UV2 region and 95\% for the UV3), the   nominal QEs (0.15 for the UV2-PMT and TPB-PMT, 0.18 for the UV3-PMT) and the solid angle estimated through the toy Monte Carlo (Sec.~\Ref{Sec:ES}). The analog manometer introduces an uncertainty of 0.2~bar on the pressure measurement. 

\begin{figure}[!t]
\centering
\includegraphics[width=0.49\textwidth]{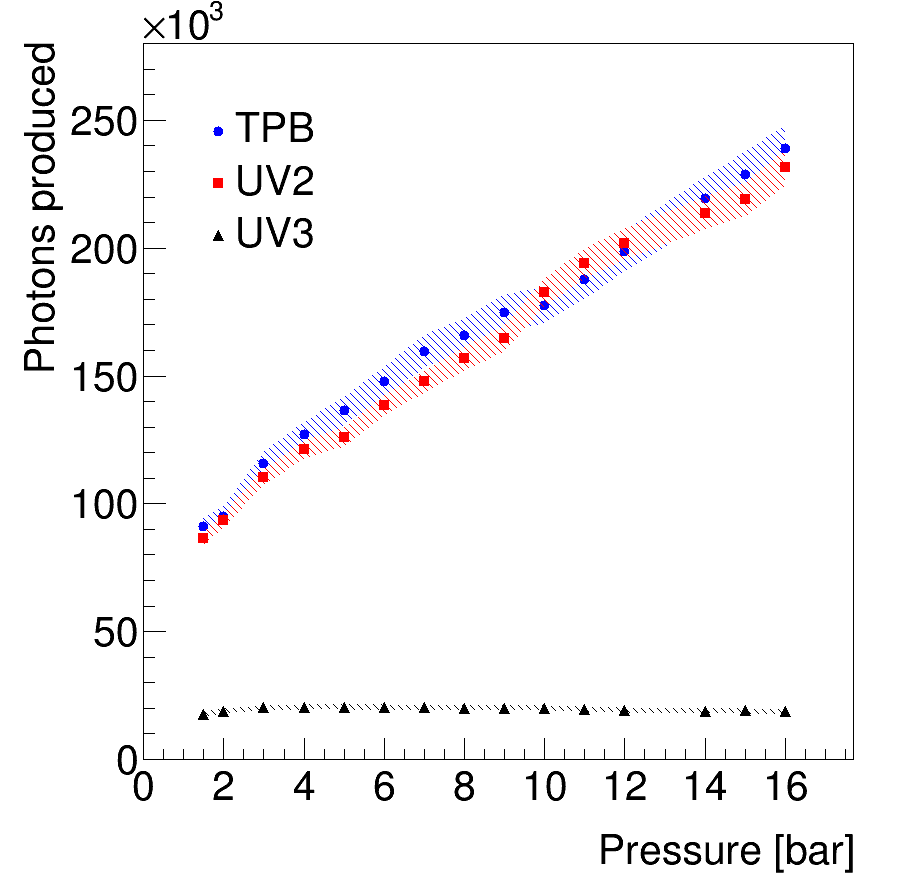}
\hspace{0.3 pt}
\includegraphics[width=0.49\textwidth]{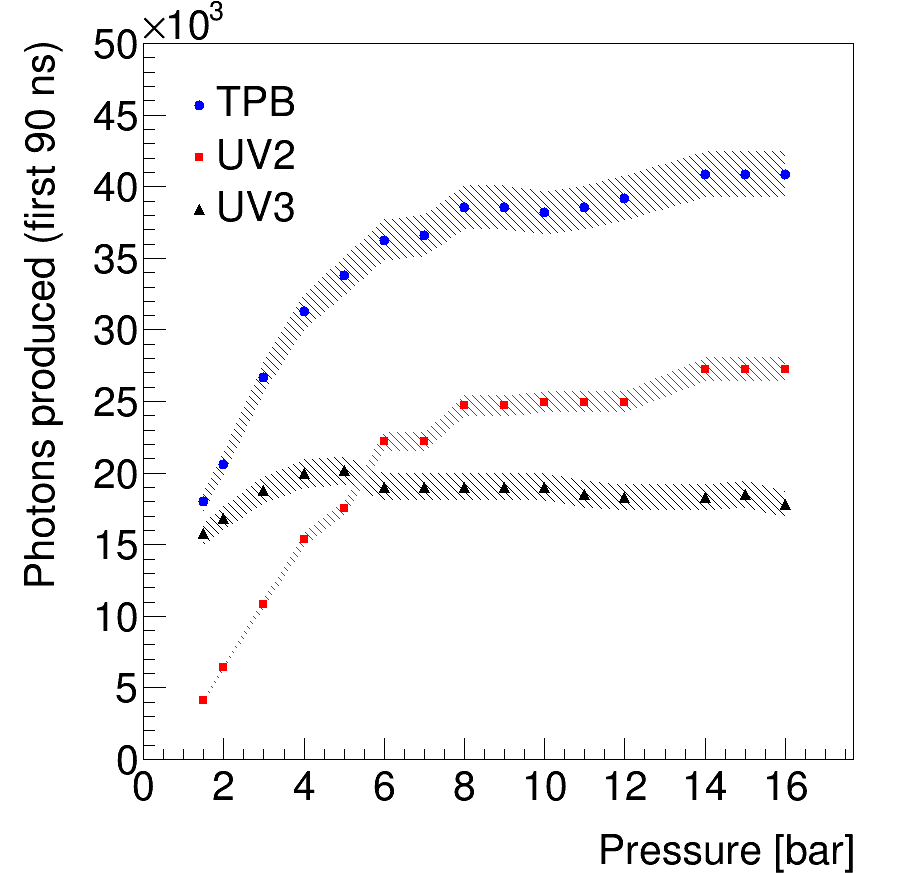}
\caption[]{(Left) Mean $\upalpha$ signal detected in the different spectral ranges as a function of the pressure. (Right) Variation with the pressure of the fast component (first 90~ns) of the argon scintillation. In both graphs, the bands indicate the systematic uncertainties. }
\label{light_evolution}
\end{figure}

The variation of the average $^{241}$Am signal, defined as the mean value of a Gaussian fit to the $\upalpha$ peak, is presented in Fig.~\ref{light_evolution}-left as a function of the gas pressure. The light yield of the chamber is sufficient to clearly identify the alpha peak in the charge spectra (Fig. \ref{source_evidence}-left), and the statistical uncertainty from the fit is negligible for all the PMTs. The fluctuations in the number of photons are taken into account with a pressure independent systematic uncertainty of $\pm~3-5$~\%, that is related to the uncertainty on the gain calibration and to the different spectral responses of the PMTs. 

The light in the UV2 region (red square in Fig.~\ref{light_evolution}-left) increases to more than twice the initial value in the [1.5,~16]~bar pressure range, an effect which is consistent with the enhanced electron-ion recombination probability at higher gas pressure~\cite{Suzuki:1981}. On the other hand, the signal in the UV3 region (black triangles) is stable within the uncertainties in that pressure range: this result suggests that the recombination light is consistent only with  128~nm photons production and no other emission is observed at longer wavelengths.

In Fig.~\ref{light_evolution}-right the evolution of the charge signal in a fixed 90~ns time window after the trigger (``fast" component) is shown as a function of the gas pressure. The UV2 signal rises up to 25~PE and gets nearly stable at pressures above 7~bar, demonstrating that more photons are emitted promptly in the UV2 region (red dots) when the gas pressure increases. The UV3 component (black dots) barely depends on the pressure and represents the dominant light emission during the first 90~ns, up to 5 bar.  

\begin{figure}[!ht]
\centering
\includegraphics[scale=0.375]{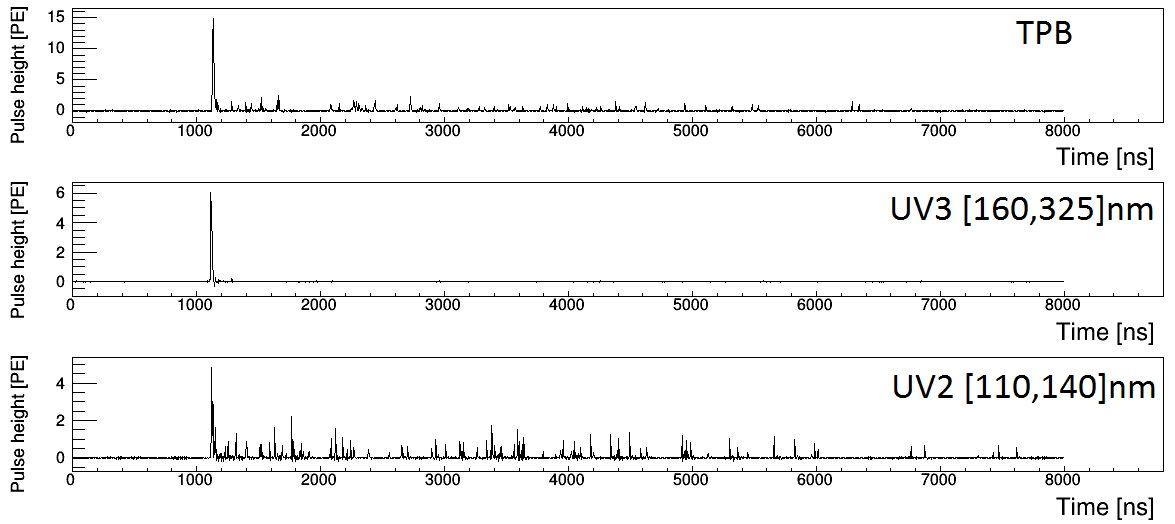}\par
\caption{Typical Ar scintillation signals from 5.5~MeV $\upalpha$-particle interaction detected in three spectral regions, for argon gas at 16~bar. Compared to the typical signal at 1.5~bar (Fig. \ref{pulse_1bar}), a fast component appears in the UV2 region when the pressure is increased.}
\label{pulse_16bar}
\end{figure}

A typical signal from an $\upalpha$ interaction in argon at 16~bar is plotted in Fig.~\ref{pulse_16bar}. When it is compared with the signal at 1.5 bar (Fig. \ref{pulse_1bar}), a fast component can be  now observed in the UV2 region. In the UV3 region, though, the pulse shape and the amplitude of the signal is similar to the 1.5 bar~case. This is consistent with the fast component of the scintillation signal detected by the TPB-PMT being significantly larger at 16~bar than at 1.5~bar. 

\begin{figure}[!ht]
\centering
\includegraphics[width=0.49\textwidth]{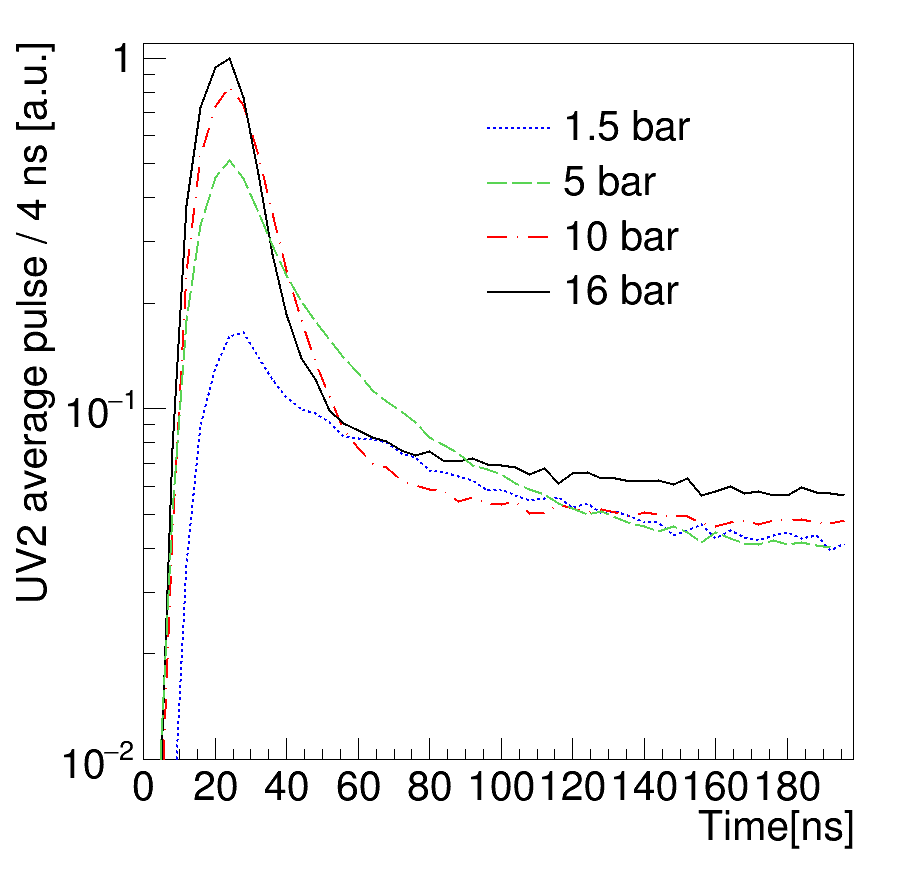}
\includegraphics[width=0.49\textwidth]{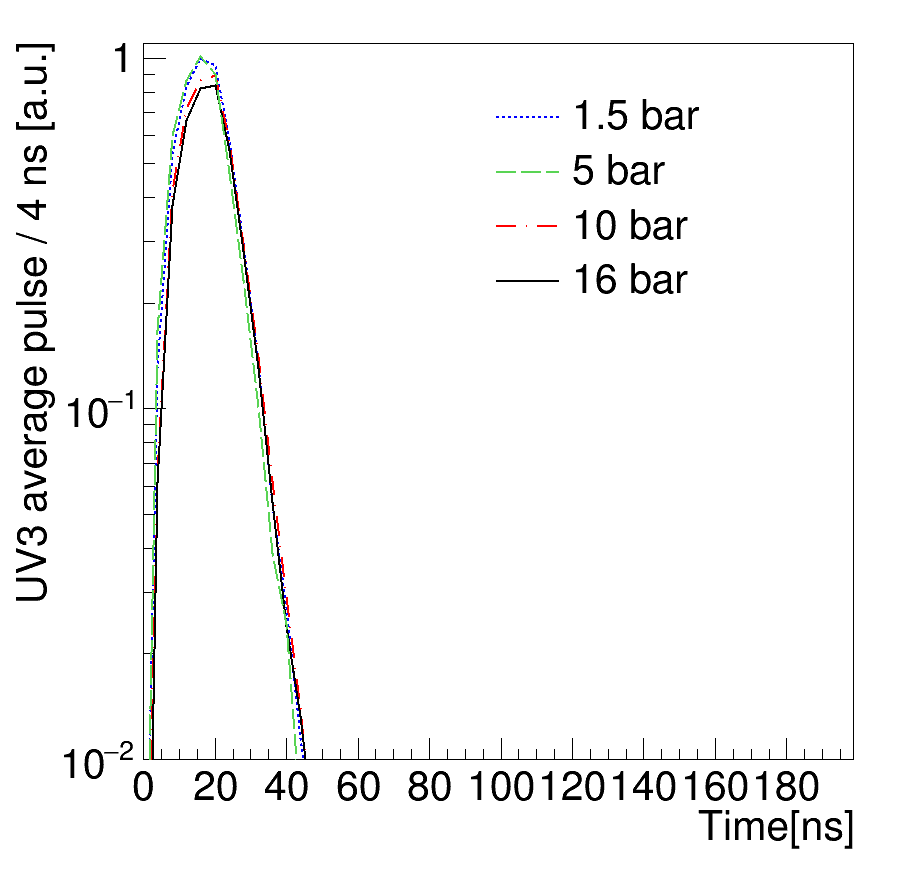}
\caption{Average signals in the [1.5,~16]~bar pressure range for the (left) UV2 and (right) UV3 regions. For UV3, only slight variations on the shapes are observed for different pressures. In the UV2 region, the fast component is more significant at higher pressure.}

\label{pressure_effect}
\end{figure}

The argon scintillation pulse, averaged over $3\times 10^4$ $\upalpha$ interactions, is depicted in Fig.~\ref{pressure_effect} for different pressures and spectral regions. The pulse heights are normalized to the maximum values, which are observed at 1.5 bar for the UV3 and 16 bar for the UV2 emissions, respectively. A relatively small decrease of the UV3 signal is evident as the pressure increases from 5 to 16~bar (Fig.~\ref{pressure_effect}-right). The enhancement of the fast argon scintillation component at higher pressure is evident in the UV2 region (Fig.~\ref{pressure_effect}-left).  This effect is consistent with the reduction of the average distance among the argon molecules at higher pressure, which allows the formation of the excimer in shorter time and leads to an increase in the fast component through the decay of the singlet $^{1}\sum_{u}$ states. 
This result is in agreement with an old study that evidenced the dependence of the Ar$^{*}_{2}$-excimer formation-time on the gas pressure~\cite{Keto:1974}. The change of slope around 6 bar in Fig. \ref{light_evolution}-right can be explained by the singlet state excimer formation time becoming smaller than the 90 ns fast signal integration window. 

\begin{figure}[!ht]
\centering
\includegraphics[width=0.49\textwidth]{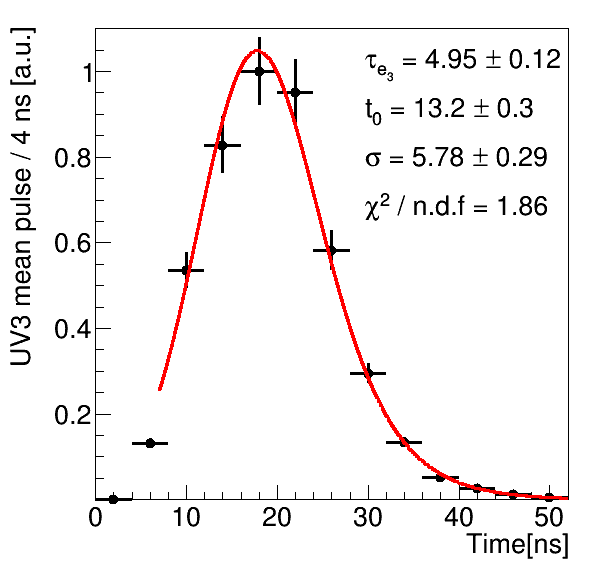}
\includegraphics[width=0.49\textwidth]{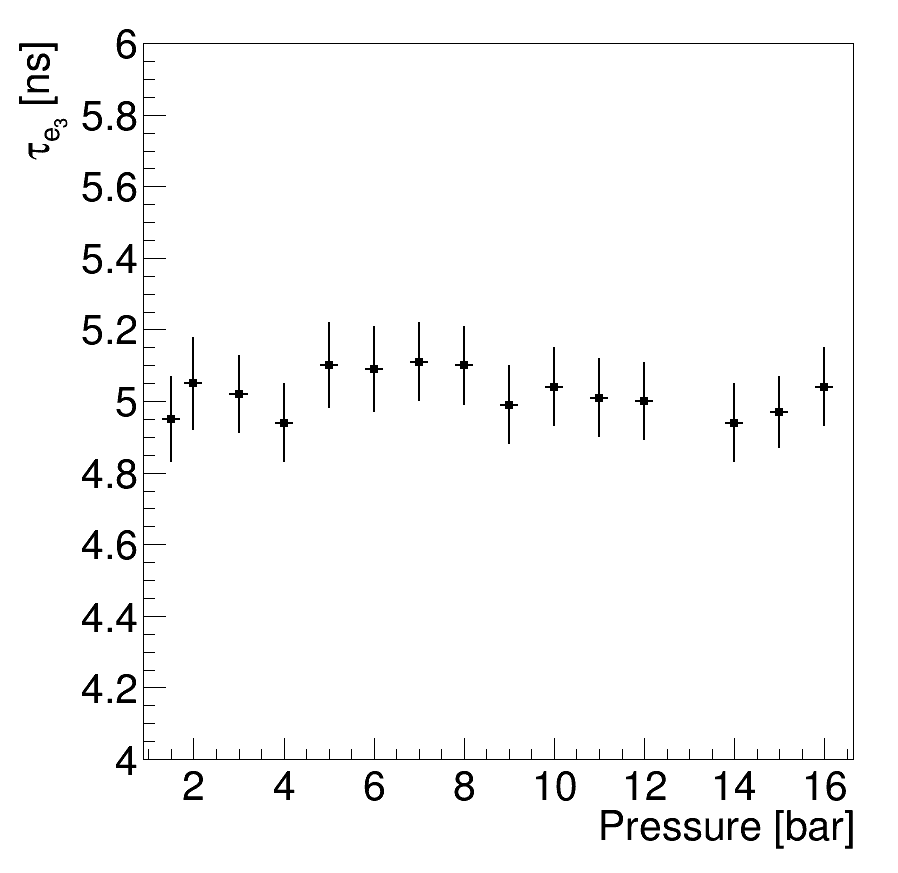}
\caption{(Left) Result of the fit (red line) of the mean pulse for $2\times 10^4$ $\upalpha$ interactions in argon at 1.5~bar in the UV3 region (black) using Eq.~\ref{2nd_eq}. (Right) Dependence of the emission time constant with pressure in the 1.5 to 16~bar range.}
\label{fit_example2}
\end{figure}

The use of sensors with different spectral sensitivity allows to study the second and the third  continuum emissions separately. The  average light pulse (from $2\times 10^4$~events) in the UV3 region  with argon at  1.5~bar is shown  in Fig.~\ref{fit_example2}-left. Data are fitted using the following expression: 

\begin{equation}
   I_{\scriptscriptstyle UV3}(t)= \frac{L}{\tau_{f_{3}}-\tau_{e_{3}}}(e^{-t/\tau_{e_{3}}}-e^{-t/\tau_{f_{3}}}) \otimes G(t-t_{0},\sigma) ,
\label{3rd_eq}
\end{equation}

\noindent where $L$ is a normalization constant, $G$ a gaussian function, with mean $t_0$ and width $\sigma$, that accounts for the detector response and $\tau_{f_{3}}$, and $\tau_{e_{3}}$ are the  time  necessary  for  the  formation of the third continuum precursors and their characteristic photon emission time. In each bin of Fig.~\ref{fit_example2}-left the pulse error is calculated through the statistical distribution of the set of waveforms and a fixed uncertainty of $\pm$~2~ns is introduced by the sampling rate of the ADC.

The $\tau_{f_{3}}$ time constant is fast and it cannot be precisely measured from the fit due to the 250~MHz maximum sampling rate of our ADC, thus a limit $\tau_{f_{3}}\lesssim 1$~ns is set on the molecular-ions formation time. The photon emission time constant $\tau_{e_{3}}$ is independent of the gas pressure in the range [1,~16]~bar (Fig.~\ref{fit_example2}-right) and its value is  5.02~$\pm$~0.11~ns, calculated as the average of the fit results in the pressure range of interest. This result is compatible with previous works~\cite{Langhoff:1994, Birot:1975}, where values of $\approx$~5~ns have been obtained for the lifetime of these radiative molecular states.

The time constants of the argon second continuum in the UV2 region are obtained in the [1.5,~8]~bar pressure range, fitting the average light pulse in the UV2 region with the following function~\cite{Amsler:2007gs}:

\begin{equation}
   I_{\scriptscriptstyle UV2}(t)= \left[\frac{L_{1}}{\tau_{f_{2}}-\tau_{e_{2}}^{s}}(e^{-t/\tau_{f_{2}}}- e^{-t/\tau_{e_{2}}^{s}})+
   \frac{L_{2}}{\tau_{f_{2}}-\tau_{e_{2}}^{t}}(e^{-t/\tau_{f_{2}}}-e^{-t/\tau_{e_{2}}^{t}})\right]
   \otimes G(t-t_{0},\sigma) ,
\label{2nd_eq}
\end{equation}

\begin{figure}[!ht]
\centering
\includegraphics[width=0.49\textwidth]{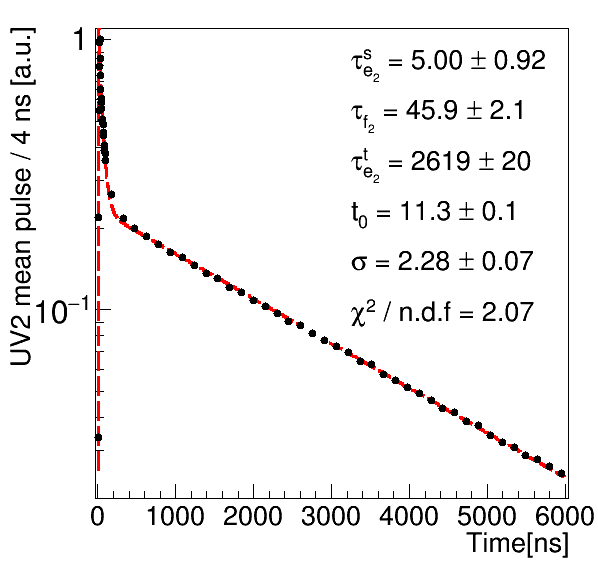}
\includegraphics[width=0.49\textwidth]{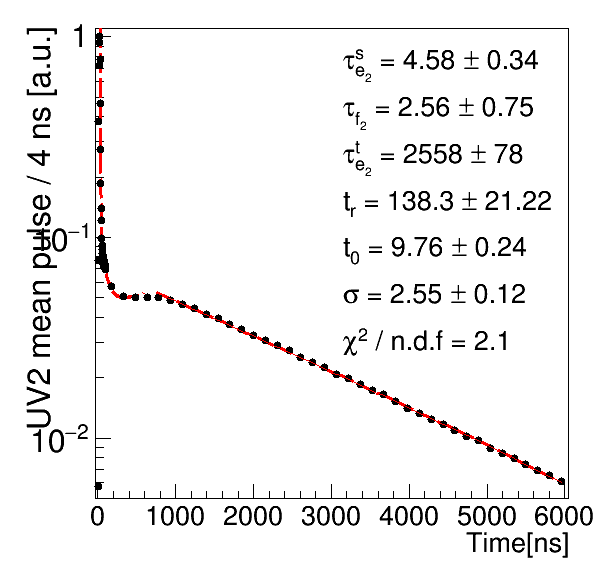}
\caption{Results of the fits (red lines) of the mean pulses for $2\times 10^4$ $\upalpha$ interactions in argon at (left) 1.5~bar and (right) 16~bar, in the UV2 region (black dots). A 4~ns binning is used up to 100~ns and it is increased to 160~ns for larger times. Eq.~\ref{2nd_eq} is used for the fit on the left, and Eq.~\ref{2nd_eq} + Eq.~\ref{reco_eq} for the fit on the right.}
\label{fit_example}
\end{figure}

The two exponential differences account for the singlet ($s$) and triplet ($t$) contributions, independently, with the corresponding $\tau_{e_{2}}$ parameters labeled accordingly. The formation times have been assumed to be identical for both contributions. The result of the fit of the UV2-PMT average charge spectrum at 1.5~bar (Fig.~\ref{fit_example}-left) proves that Eq.~\ref{2nd_eq} represents an accurate description of the argon scintillation signal at that pressure.

\noindent At pressures larger than 8~bar, the UV2 signal shape is characterized by a new structure  a few $\upmu$s after the maximum pulse height, and Eq.~\ref{2nd_eq} no longer describes the spectrum. A tentative explanation of this feature is taken from the model proposed in ~\cite{Kubota:1979}, that addresses the electron-ion recombination luminescence in the absence of  electric field. This model considers a region of uniform ionization density and neglects the diffusion process out of the $\upalpha$ track for the thermalized electrons. 
The recombination time $\tau_r$ is proportional to $P^{-2.7\pm0.3}$  for pressures larger than 10~bar ~\cite{Suzuki:1981} and it becomes shorter at larger pressures. In liquid argon, this time is below 1~ns, where the time dependence of the recombination is dominated by the molecular de-excitation time.

\begin{figure}[!ht]
\centering
\includegraphics[width=0.49\textwidth]{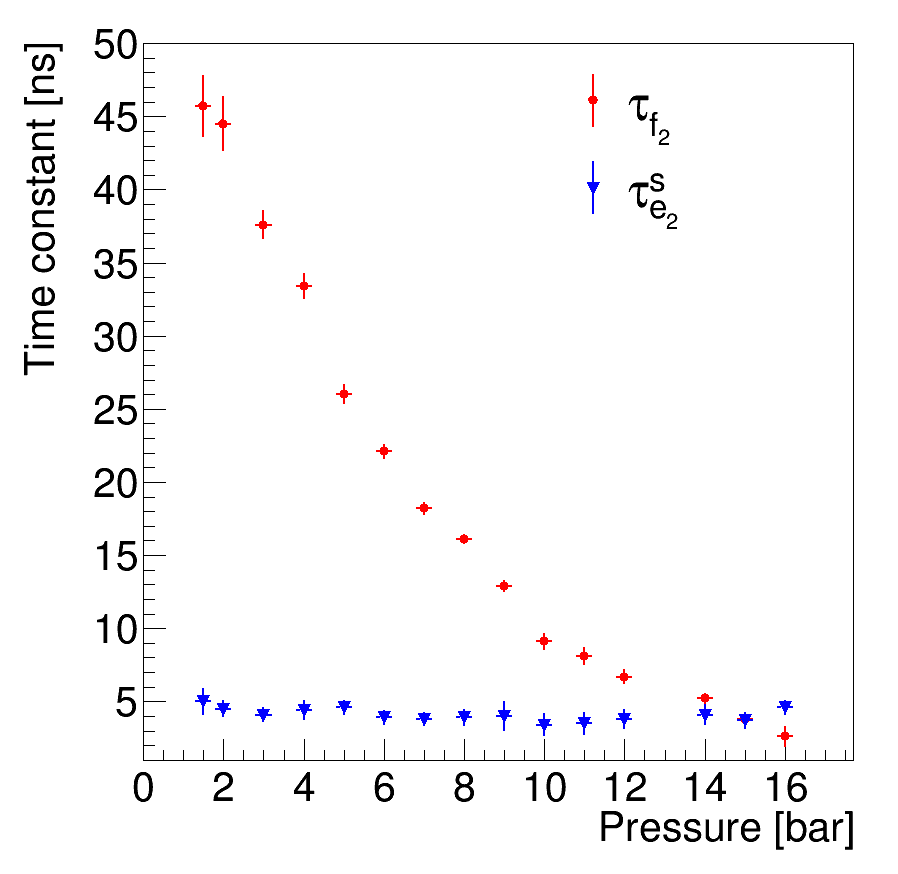}
\includegraphics[width=0.49\textwidth]{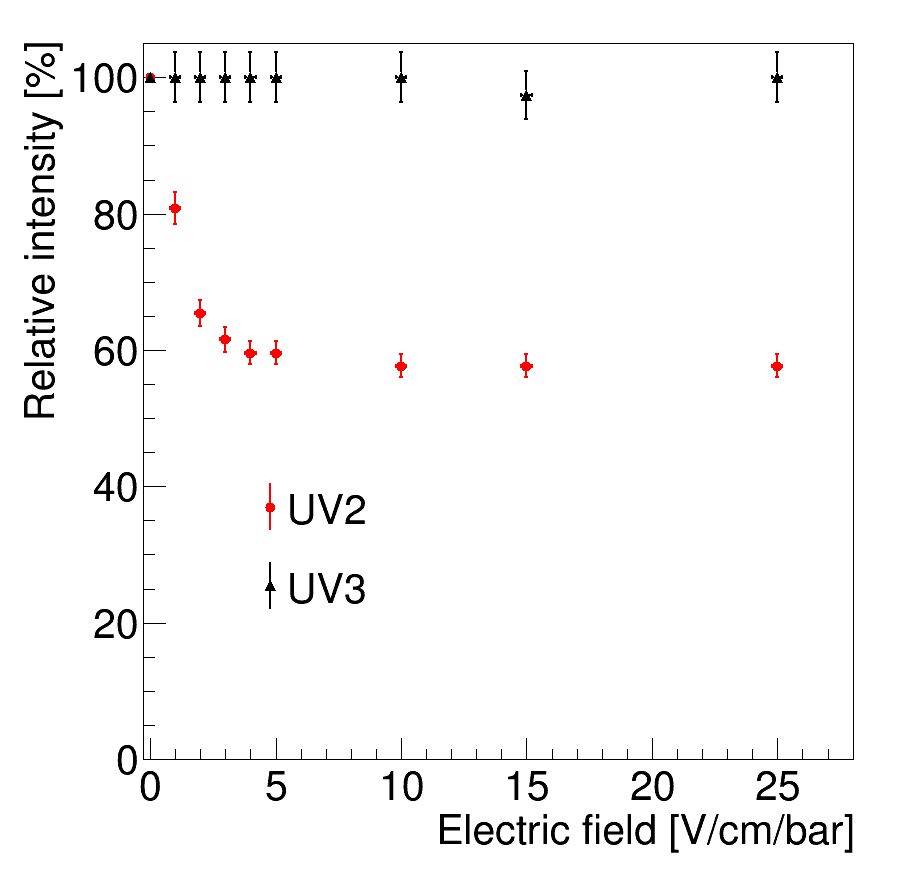}
\caption{(Left) Dependence of the excimer  formation time constant (red dots) and singlet de-excitation lifetime (blue triangles) with pressure in the 1.5 to 16~bar range. (Right) Dependence of the luminescence with the electric field intensity in the UV2 (red dots) and UV3 (black triangles) regions. The UV3 signal is not affected by the electric field. }
\label{second_constant}
\end{figure}

A new term is added to Eq.~\ref{2nd_eq} in order to describe the scintillation from the charge recombination, $I_{\scriptscriptstyle UV2}^{\scriptscriptstyle R}$, which depends on the characteristic recombination time, $t_r$, and the excimer lifetime, $\tau_e$, (discussed in Sec. \ref{Sec:sci}): 

\begin{equation}
\begin{split}
   I_{\scriptscriptstyle UV2}^{\scriptscriptstyle R}(t) & = \; L_{3} \; ( e^{-t/\tau_{e}^{t}}-e^{-2t/t_{r}} ), \qquad  \textnormal{for} \; t<t_{r} , \\
   I_{\scriptscriptstyle UV2}^{\scriptscriptstyle R}(t) & = \; L_{3} \; e^{-t/\tau_{e}^{t}}, \hspace{19mm}\qquad \textnormal{for} \; t>t_{r} ,
\end{split}
\label{reco_eq}
\end{equation}

\noindent where $L_3$ is a normalization constant. The result of the UV2 signal fit to Eq.~\ref{2nd_eq}, after including the additional term $I_{\scriptscriptstyle UV2}^{\scriptscriptstyle R}(t)$, is shown in Fig.~\ref{fit_example}-right. The signal is successfully described by the fit above 8~bar up to 16~bar. The dependence of the excimer formation time  and the singlet decay time in the UV2 region are summarized in Fig.~\ref{second_constant}-left for different pressures.

The excimer formation time has a strong pressure dependence, decreasing from 46~ns at 1.5~bar to 2.6~ns at 16~bar. On the other hand, the singlet decay time emission is independent of the pressure with values around 4-5~ns. The triplet lifetime is measured to be $\approx$~3 $\upmu$s  depending on the gas flow as expected. These results are in good agreement with previous studies~\cite{Keto:1974}. At argon pressures below 5 bar, the  excimer formation  time is the dominant factor that determines the photon emission during the first hundred of ns. In this case, the singlet component of the second continuum emission is smeared over tens of ns. At higher gas pressure, the typical excimer formation time decreases. 

\section{Electric field dependence of the scintillation and spectroscopic studies of the electron-ion recombination}
\label{Sec:ResE}

With the aim of establishing the nature of the new structure that appears in the UV2 range at large pressures and to study the field dependence of the UV3 emission, a small field cage (2~cm height) with the $\rm ^{241}Am$ source on the anode plate was introduced in the central volume of the pressure chamber.

Data at 10~bar of pressure were taken in order to fully contain the $\upalpha$ track in the field region. In this setup the anode is grounded and the reduced electric field (E/P) applied in the range from 0 to 25~V/cm/bar. The number of PE measured with the $\rm ^{241}Am$ peak as a function of the electric field is shown in Fig.~\ref{second_constant}-right, evidencing that the emission in the UV2 wavelength range decreases with electric fields up to 4 V/cm/bar and then remains constant. The saturation in the collection of the charge above this field value is consistent with the measurements reported in~\cite{Suzuki:1981} for the second continuum.  On the other hand, the emission in the UV3 spectral region is not affected by the electric field up to 25~V/cm/bar.

\begin{figure}[!t]
\centering
\includegraphics[width=0.49\textwidth]{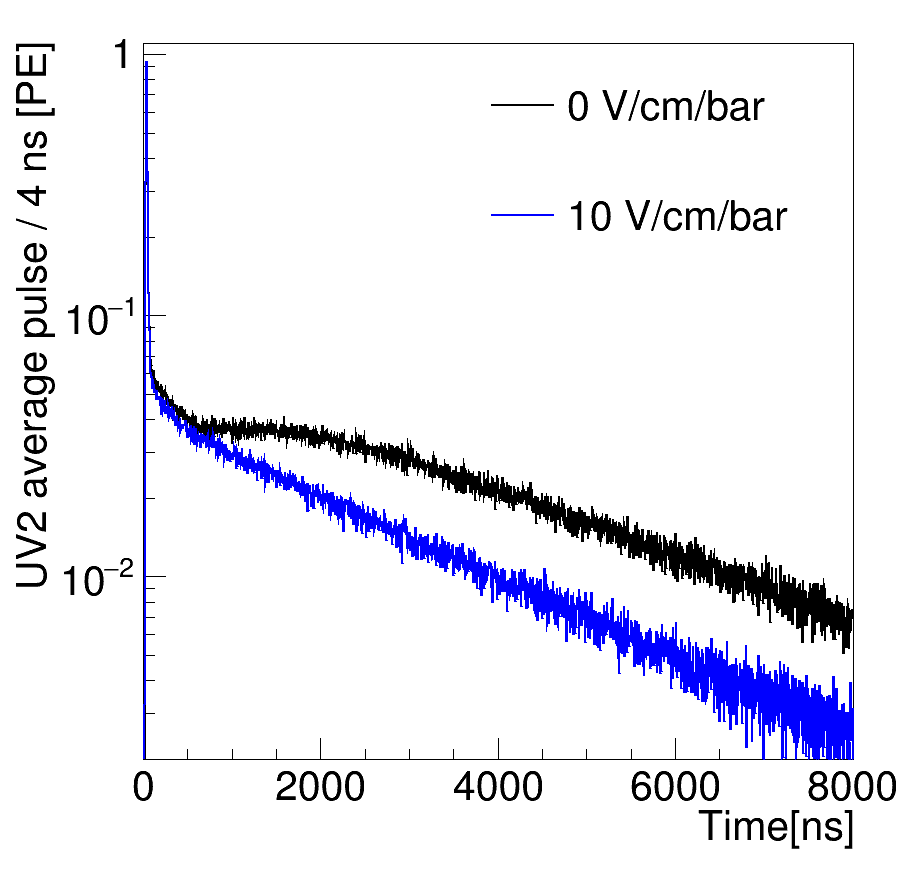}
\includegraphics[width=0.49\textwidth]{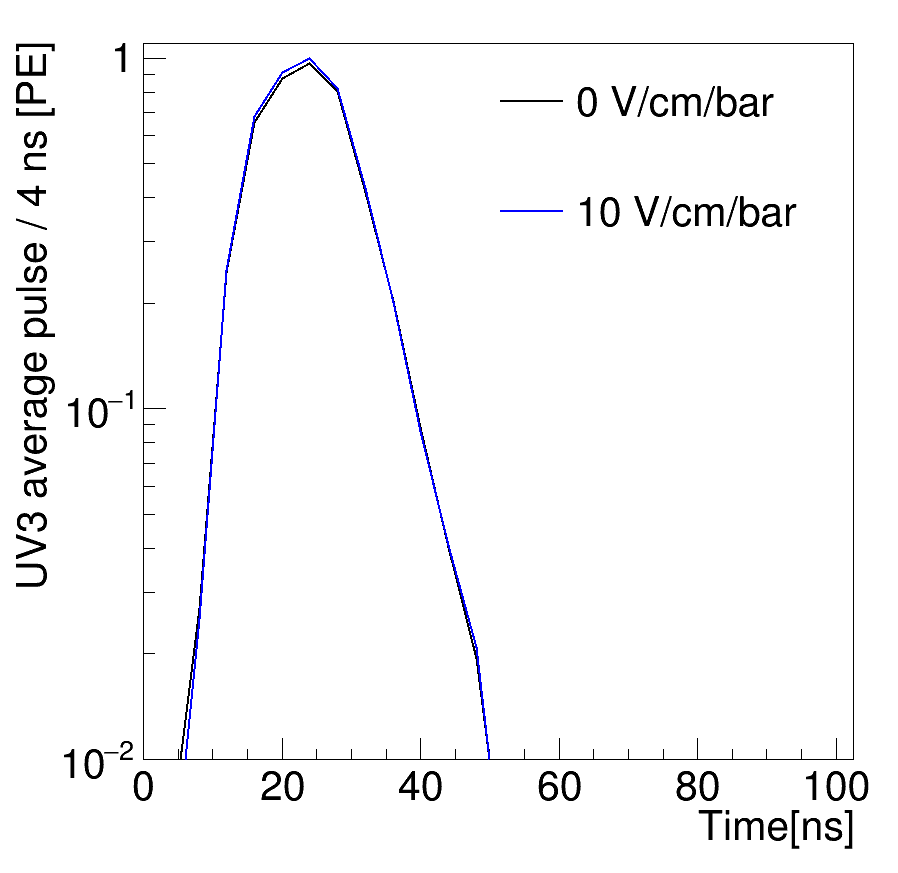}
\caption{Mean pulse shape of the  UV2  (left) and  UV3 (right) signals with (blue line) and without (black line) electric field with argon at 10 bar. The bump in the UV2 graph disappears when the electric field is applied. The UV3 signal is not affected by the electric field.}
\label{reco_effect}
\end{figure}

The scintillation pulses at 10~bar with and without a 100~V/cm field, averaged over $2\times 10^4$ events and normalized to the maximum value of the distribution,  are shown in Fig.~\ref{reco_effect}, both for the UV2 and UV3 spectral ranges. The bump over 1 $\upmu$s in the UV2 range (left graph) disappears by increasing the amplitude of the reduced field. This result confirms the interpretation of this structure in terms of electron-ion recombination given in Sec. \ref{Sec:ResP}. When a sufficiently strong field is applied, the charge recombination is suppressed and the overall signal shape is similar to the scintillation pulse  at pressures below 8~bar. In these conditions, the argon slow component is well described by a single exponential  function.  

The distribution for UV3 (right plot) proves that the photon emission in this spectral region is not affected by the electric field. This result demonstrates that the recombination light is consistent only with the second continuum emission at 128~nm.

\section{First evidence of the third continuum emission with a 
\texorpdfstring{$\mathbf{\beta}$}{beta} source}

Our experiments with $\upalpha$ particles in argon at different pressures have proved that a substantial component of the scintillation is in the range [160,~325]~nm and is not related to the radiative de-excitation of the singlet and triplet excimers. We interpret this photon production  through the mechanism responsible for the third continuum emission (Sec.\ref{Sec:sci}).
Additional studies have been performed  replacing the $\rm ^{241}Am$ $\upalpha$ source with a weak $\rm ^{90}Sr/Y$ $\upbeta$ source (activity $\approx$~100~Bq), with  546~keV and 2280~keV  Q-values of the decays. We collected several runs with argon in the pressure range [15,~21]~bar. The $\upbeta$'s from the source are typically not fully contained in our detector, unlike for the case of $\upalpha$ interactions.
To increase the light collection, the UV2-PMT (R6835) is replaced by another phototube coated  with TPB (R6095). In this configuration, the trigger is produced by the coincidence of the two TPB-PMTs. 

\begin{figure}[!ht]
\centering
\includegraphics[width=0.6\textwidth]{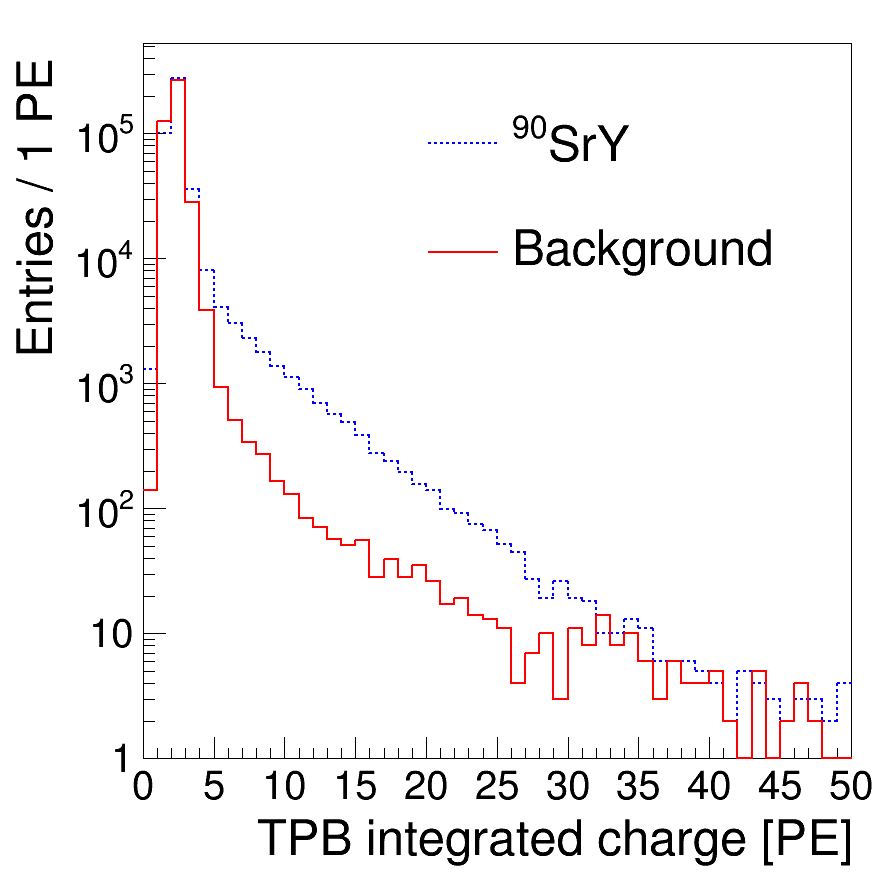}\par
\caption{Integrated charge for events collected without (red line) and with (blue line) the $\rm ^{90}Sr/Y$ $\upbeta$ source in the center of the detector. The number of events is normalized to the same acquisition time.}
\label{beta_source}
\end{figure}

\begin{figure}[!ht]
 \centering
\includegraphics[scale=0.37]{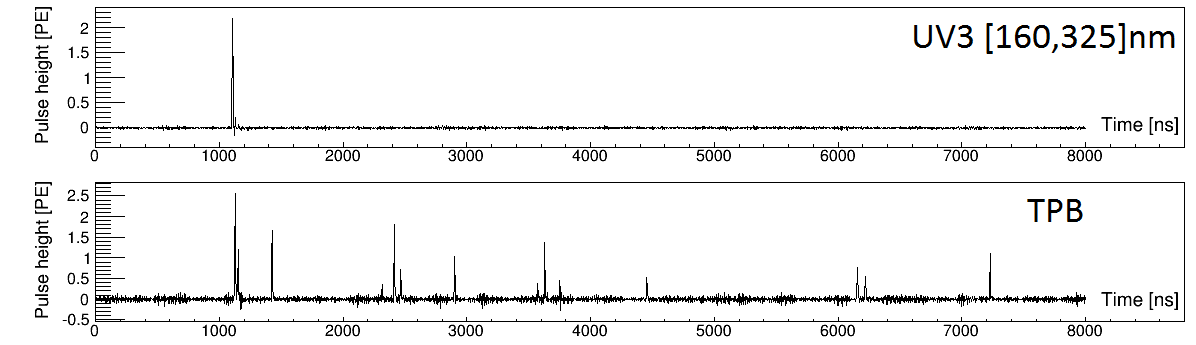}\par
\caption{Typical Ar scintillation signals detected for a $\upbeta$ event in two spectral regions, with argon gas at 20~bar.}
\label{beta_mean}
\end{figure}

The integrated charge spectrum of one TPB-PMT, obtained with the $\rm ^{90}Sr/Y$ $\upbeta$ source and gas argon at 20~bar is compared with a background spectrum taken without any source (Fig.~\ref{beta_source}). A prominent excess is present at low energy, between 6 and 25~PE, giving us a solid evidence of the actual identification of the $\upbeta$ interactions. A typical signal produced by one UV-PMT and one TPB-PMT in this region is displayed in Fig.~\ref{beta_mean}. A clear pulse,  similar to the one detected with the $\upalpha$ source but with an amplitude consistent with the different energy scale, is shown in the UV3 region in coincidence with the TPB-PMT signal. To the best of the authors knowledge, this result  is the first evidence of the third continuum emission produced by a $\upbeta$ interaction in argon gas. 

This demonstration of the third continuum emission from $\upalpha$ and $\upbeta$  interactions in argon, provides evidence that the temporal and spectroscopic information of the argon scintillation are strongly related, and can be used independently or complementary for particle identification. New studies focused on a novel particle identification technique based on the spectroscopic analysis of the light pulses are currently on-going \cite{Santorelli:PID}. 

\section{Conclusions}

Conventional dark matter and neutrino detectors based on noble element time-projection chambers are built with fast light detection devices coupled with photon wavelength-shifters that integrate the light signal over a wide spectral range, missing the potential information provided by the scintillation wavelength.  

We have studied the primary and the recombination scintillation  of the argon gas with a wavelength sensitive detector operated with  $\upalpha$ and $\upbeta$ sources  electrodeposited on  stainless steel disks. Our results evidence the emission of photons at  wavelengths larger than the 128~nm line through a  production mechanism which is not based on the excimer formation associated to the low-lying atomic states. We interpret this component of the argon scintillation  as the third continuum emission.

We have proven that up to 20\% of the scintillation obtained with an $\rm ^{241}$Am source in 1.5~bar argon gas is in the range $[160,~325]$~nm. The photon yield and the typical emission time of this  component  are largely independent from the gas pressure up to 16~bar, and the emission is not significantly affected by an external electric field.  Compared to the second continuum, which is   dominated by the excimer formation time, the third continuum is remarkably fast and represents the main contribution to the argon light signal during the first tens of ns, for pressures below 10 bar. Evidence of the third continuum emission produced by $\upbeta$ interactions in argon gas has been also obtained for the first time using a $\rm ^{90}Sr/Y$  source. The spectroscopic studies of the  electron-ion recombination light revealed that this component is consistent  with the 128~nm emission.

We have established that the argon second and the third continuum scintillation can be distinguished experimentally by means of sensors with different spectral sensitivities, making possible to exploit distinctive features of the noble gases photon emission that are not envisaged by the present experiments. Particularly, our investigations open new paths toward a novel particle identification technique based on the spectral information of the noble-elements scintillation light.

\section*{Acknowledgments}
This research is funded by the Spanish Ministry of Economy and Competitiveness (MINECO) through the grant FPA2017-92505-EXP. The authors are also supported by the ``F\'{i}sica de part\'{i}culas" Unit of CIEMAT through the grant MDM-2015-0509. DGD is supported by the Ramon y Cajal program (Spain) under contract number RYC-2015-18820.


\end{document}